\documentclass[prl, showpacs, twocolumn, nofootinbib, superscriptaddress, amsmath, amssymb, floatfix]{revtex4-1}
\usepackage{pdfpages}
\usepackage{amsmath,amssymb,bm}
\usepackage{graphics}
\usepackage{epstopdf}
\usepackage{epsfig}
\usepackage{graphicx}
\usepackage{eqnarray,amsmath}
\usepackage{color}
\usepackage[normalem]{ulem} 
\usepackage{graphicx}
\usepackage{dcolumn} 
\usepackage{bm}      
\usepackage{subfigure}
\usepackage{amsfonts}
\usepackage{rotating}
\usepackage{fontenc}
\usepackage{cancel}
\usepackage{amsthm}
\usepackage{hyperref}

\makeatletter
\AtBeginDocument{\let\LS@rot\@undefined}
\makeatother

\newcommand{\be}{\begin{equation}}
\newcommand{\ee}{\end{equation}}
\newcommand{\bea}{\begin{eqnarray}}
\newcommand{\eea}{\end{eqnarray}}
\newcommand{\bw}{\begin{widetext}}
\newcommand{\ew}{\end{widetext}}
\newcommand{\mm}{\mathrm}

\usepackage{xcolor}

\newcommand{\bi}{\begin{itemize}}
\newcommand{\ei}{\end{itemize}}

\begin{document}
\title{Covering Problems and Core Percolations on Hypergraphs}

\author{Bruno Coelho Coutinho}
\email{bruno.c.coutinho@gmail.com}
\affiliation{Instituto de Telecomunica\c{c}\~{o}es, Physics of Information and Quantum Technologies Group, Portugal}

\author{Hai-Jun Zhou}
\affiliation{CAS Key Laboratory for Theoretical Physics, Institute of Theoretical Physics, Chinese Academy of Sciences, Beijing 100190, China}

\author{Yang-Yu Liu}
\email{yyl@channing.harvard.edu}
\affiliation{Channing Division of Network Medicine, Brigham and
  Women's Hospital and Harvard Medical School, Boston, Massachusetts 02115, USA} 
\affiliation{Center for Cancer Systems Biology, Dana-Farber Cancer
  Institute, Boston, Massachusetts 02115, USA}

\begin{abstract}
We introduce two generalizations of core percolation in graphs to
hypergraphs, related to the minimum hyperedge cover problem and the
minimum vertex cover problem on hypergraphs, respectively. We offer
analytical solutions of these two core percolations for random
hypergraphs with arbitrary vertex degree and hyperedge cardinality
distributions. We find that for several real-world hypergraphs their
two cores tend to be much smaller than those of their randomized
counterparts, suggesting that covering problems in those real-world
hypergraphs can actually be solved in polynomial time. 
\end{abstract}

\maketitle


As a natural generalization of a graph, a hypergraph consists of
vertices and hyperedges ~\cite{Bretto:2013:HTI:2500991}. A hyperedge
can simultaneously connect any number of vertices, which facilitates a
more faithful representation of many real-world
networks~\cite{PhysRevE.93.062311,PhysRevE.93.032315}. 
For example, given a set of proteins and a set of protein complexes,
the corresponding hypergraph naturally captures the information on
proteins that interact within a protein complex~\cite{klamt2009hypergraphs}. 
%
%
For a biochemical reaction system, the hypergraph representation 
indicates which biomolecules participate in a particular
reaction~\cite{klamt2009hypergraphs,1303205}. 
In computer science, the factorization of complicated global functions of many variables can
often be represented by a factor graph, wich can be mapped to a
hypergraph~\cite{Kschischang-etal-2001,klamt2009hypergraphs,klamt2009hypergraphs,1303205,Academicteam_hypergraphs,Motifs_coauthorship,PhysRevE.93.032315}. 
In social science, a collaboration network can also be represented by
a hypergraph, where vertices represent individuals and hyperedges connect
individuals who were involved in a specific collaboration, e.g., a
scientific paper, a patent, a consulting task, or an art
performance~\cite{Academicteam_hypergraphs,Motifs_coauthorship}. 
%
%

As in graphs, the degree of a vertex in a hypergraph is the number of
hyperedges that connect to it.
The number of vertices connected by a hyperedge is called the cardinality of
that hyperedge. 
If all hyperedges have the same cardinality $K$, the hypergraph is
said to be uniform or $K$-uniform. Note that a graph is just a
2-uniform hypergraph. 
%
%
%
%

The core of a graph --- defined as the remainder of the greedy leaf
removal (GLR) procedure where leaves (vertices of degree one) and
their neighbors are removed iteratively from the graph --- has been
related to the conductor-insulator
transition~\cite{PhysRevLett.86.2621}, structural
controllability~\cite{liu11}, and many combinatorial optimization
problems~\cite{4568355}. 
Indeed, the core size is related to a fundamental combinatorial
problem --- the minimum vertex cover (MVC) problem, which aims to find
the smallest set of vertices in a graph so that every edge is incident
to at least one vertex in the set~\cite{PhysRevLett.84.6118}. If the
core is absent, then the MVC problem is solvable in
polynomial. Otherwise, if the core exists and is extensive in
size, then the MVC problem is generally
NP-hard~\cite{PhysRevLett.84.6118,0305-4470-36-43-028}. 
As the dual of the MVC problem, the minimum edge cover problem aims to
find the smallest set of edges so that for every vertex in the graph
there is at least one edge incident to it. 
Both covering problems can be defined similarly on hypergraphs. 
The minimum edge cover problem on graphs can be computed in polynomial
time~\cite{garey1979computers}. Yet, this is not true for hypergraphs,
where both the minimum hyperedge cover (MHC) and the MVC problems are
generally NP-hard~\cite{Covering_edges_thesis}.  
%
%
Note that the MVC and MHC on hypergraphs are related to many real-world
problems, e.g., finding the optimal drug combination in
pharmacology~\cite{Vazquez2009}, searching files in a storage
systems~\cite{1053689}, etc.  
Typically these problems can be solved using approximate algorithms,
e.g., highest-degree-first~\cite{Vazquez2009,JOHNSON1974256} and
simulated annealing~\cite{Vazquez2009}. 
Here we show that those approximate algorithms are not always
necessary. To achieve that, we extend the concept of the core in
graph to the hypergraph case, and define two cores associated with the
MVC and MHC problem, respectively.   

\begin{figure}[t!]
\centering
\includegraphics[width=0.45\textwidth]{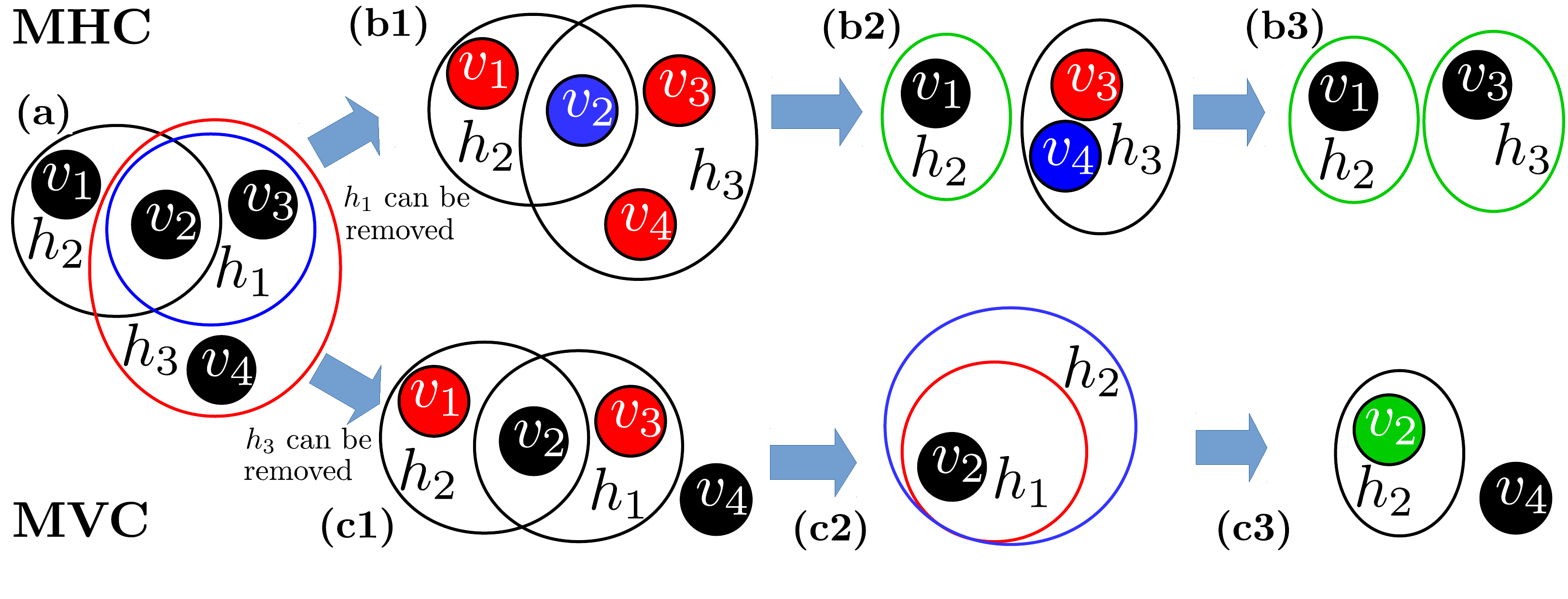}
\caption{Generalized greedy leaf removal helps us solve the minimum
  hyperedge cover (MHC) and the minimum vertex cover (MVC) problems on
  hypergraphs. Vertices are represented by dots and hyperedges by
  circles. The green vertices and hyperedges in (b3) and (c3) are the
  solution to the MHC or MVC problem, respectively.
  \label{gglr}} 
\end{figure}

Let us consider the MHC problem of the hypergraph  $\mathcal{H}_0$ in
Fig.~\ref{gglr}(a), which has three hyperedges $\{h_1, h_2, h_3\}$ and
four vertices $\{v_1, v_2, v_3, v_4\}$. The hyperedge   $h_3$ contains
all the vertices in  $h_1$, as well as vertex $v_4$, thus $h_1$ is not
necessary for the MHC and can be removed, leading to the hypergraph
$\mathcal{H}_{1}$ shown in Fig.~\ref{gglr}(b1). In $\mathcal{H}_{1}$
the vertex $v_2$ is contained by hyperedge $h_2$ that contains also
vertex $v_1$. Hence if $v_1$ is covered,  $v_2$ is also covered.  We
can therefore remove $v_2$ from $\mathcal{H}_{1}$. By iteratively
removing  vertices and hyperedgse using these rules, we get the
hypergraph shown  in Fig.~\ref{gglr}(b3), for which solving the MHC
problem is trivial.  
The MVC problem is dual to the MHC problem, hence we can obtain a dual
set of rules. In the hypergraph $\mathcal{H}_0$, the hyperedge  $h_3$ contains all the vertices in
$h_1$, as well as vertex $v_4$, thus if $h_1$ is covered $h_3$ is also
covered. We can therefore remove $h_3$ and  obtain the hypergraph
$\mathcal{H}_2$ in Fig.~\ref{gglr}(c1), for which we find that vertex
$v_1$ is redundant and can be removed (since $v_2$ covers the same
hyperedge  as $v_1$ and also covers hyperedge $h_1$). By iteratively
removing vertices and hyperedgse using these rules, we get the
hypergraph shown in Fig.~\ref{gglr}(c3), for which the MVC problem is
trivial. 

The example shown in Fig.~\ref{gglr} prompts us to define three sets
of hyperedges (or vertices): (1) $S$ is a solution of the MHC (MVC)
problem; (2) $\tilde{S}$ contains hyperedges (vertices) that can be
determined to be part of $S$ using our approach; and (3) the core
h$_{\rm core}$ (v$_{\rm core}$) contains the hyperedges (vertices)
that cannot be determined if they belong to $S$ or not using our
approach. 
If the degree (cardinality) of a vertex (hyperedge)
is zero, then it can not be covered.  For the 
remaining vertices (hyperedges) we can find the vertices (hyperedges)
that belong to each category based on the following rules. 
\emph{(Rule-1)}: Consider hyperedges $h_1$ and $h_2$ that contain the set of vertices
$V_1$ and $V_2$. If $V_1$ and $V_2$ are not empty sets, and $V_1
\subseteq  V_2$, we remove $h_1$ (or $h_2$) to solve the MHC (or MVC)
problem, respectively. 
\emph{(Rule-2)}: Consider vertices $v_1$ and $v_2$ that are contained
by the set of hyperedges $H_1$ and $H_2$.  If $H_1$ and $H_2$ are are
not empty sets, and  $H_1 \subseteq  H_2$, we remove $v_2$ (or $v_1$)
to solve the MHC (or MVC) problem, respectively. 
%
We repeat this process until no more vertices or hyperedges can be
removed. In the final hypergraph, hyperedges (vertices) with
cardinalty (degree) one belong to $\tilde{S}$ (Fig.~\ref{gglr}(b3)
and (c3)); hyperedges (vertices) with cardinalty (degree) larger
than one belong to h$_{\mm{core}}$ (v$_{\mm{core}}$).   

We emphasize that these two rules can be considered as the generalized GLR
procedure on hypergraphs, which reduces to the standard GLR procedure
on graphs. Note that even if the resulting h$_{\mm{core}}$
(v$_{\mm{core}}$) is very small but non-zero, the generalized GLR
procedure is better than approximation algorithms in solving the MHC
(MVC) problem, because it explicitly tells us which hyperedges (vertices) belong
to the solution $S$, which do not, and which cannot be determined.

\begin{figure}[t!]
\centering
\includegraphics[width=0.45\textwidth]{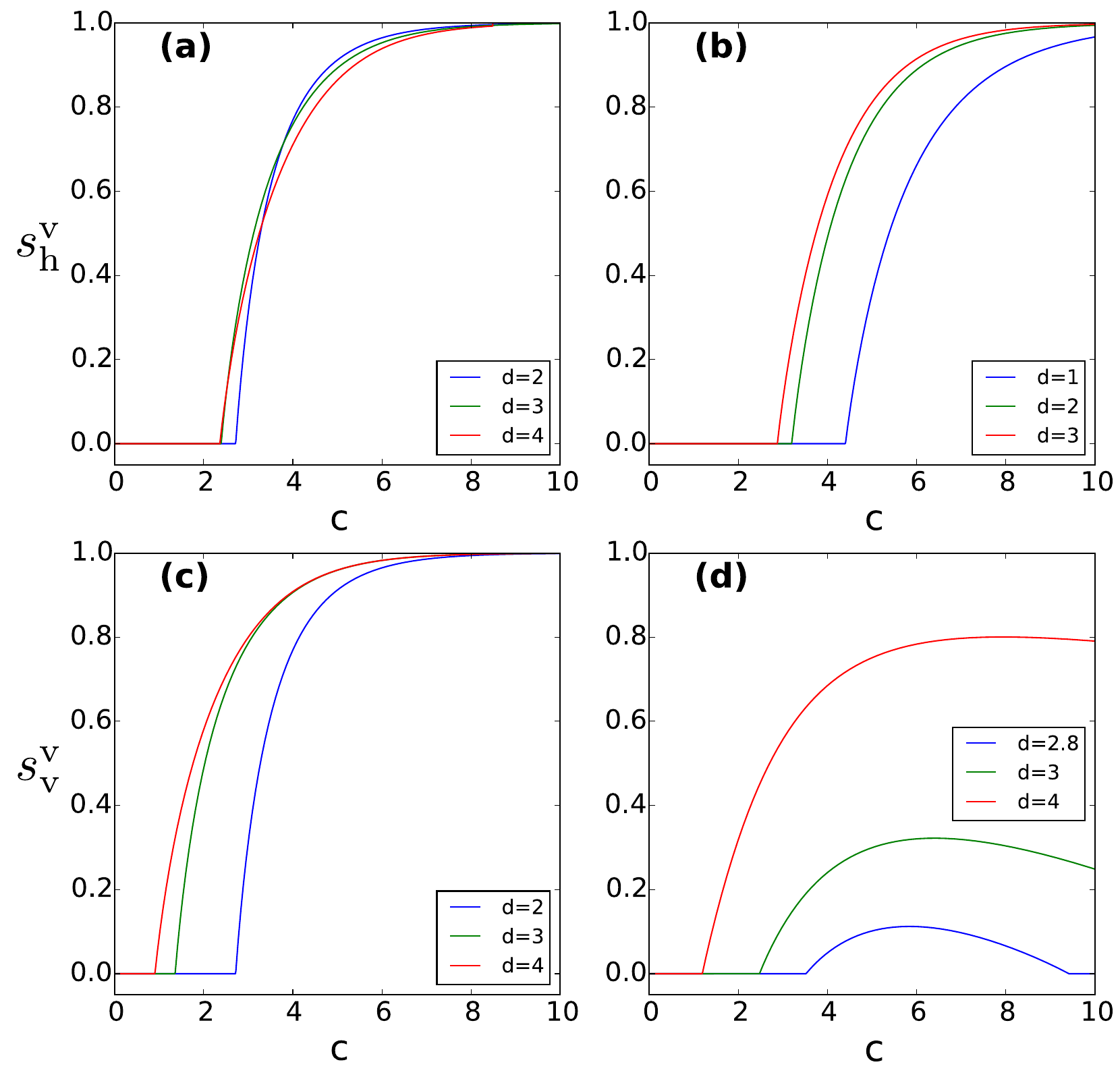}
\caption{(Color online) The relative size of the core size hypergraphs
  with Poisson vertex degree distributions, (a,b) $s^\text{v}_\text{h}$. (c,d) $s^\text{v}_\text{v}$. In (a) and (c) we consider $d$-uniform hypergraphs, meaning that all hyperdege have the same cardinality $d$. In (b) and (d)  we consider that the cardinality of the hyperdeges follows a Poisson distribution with average cardinality $d$.
\label{graph3}}
\end{figure}

\begin{figure}[t!]
\centering
\includegraphics[width=0.45\textwidth]{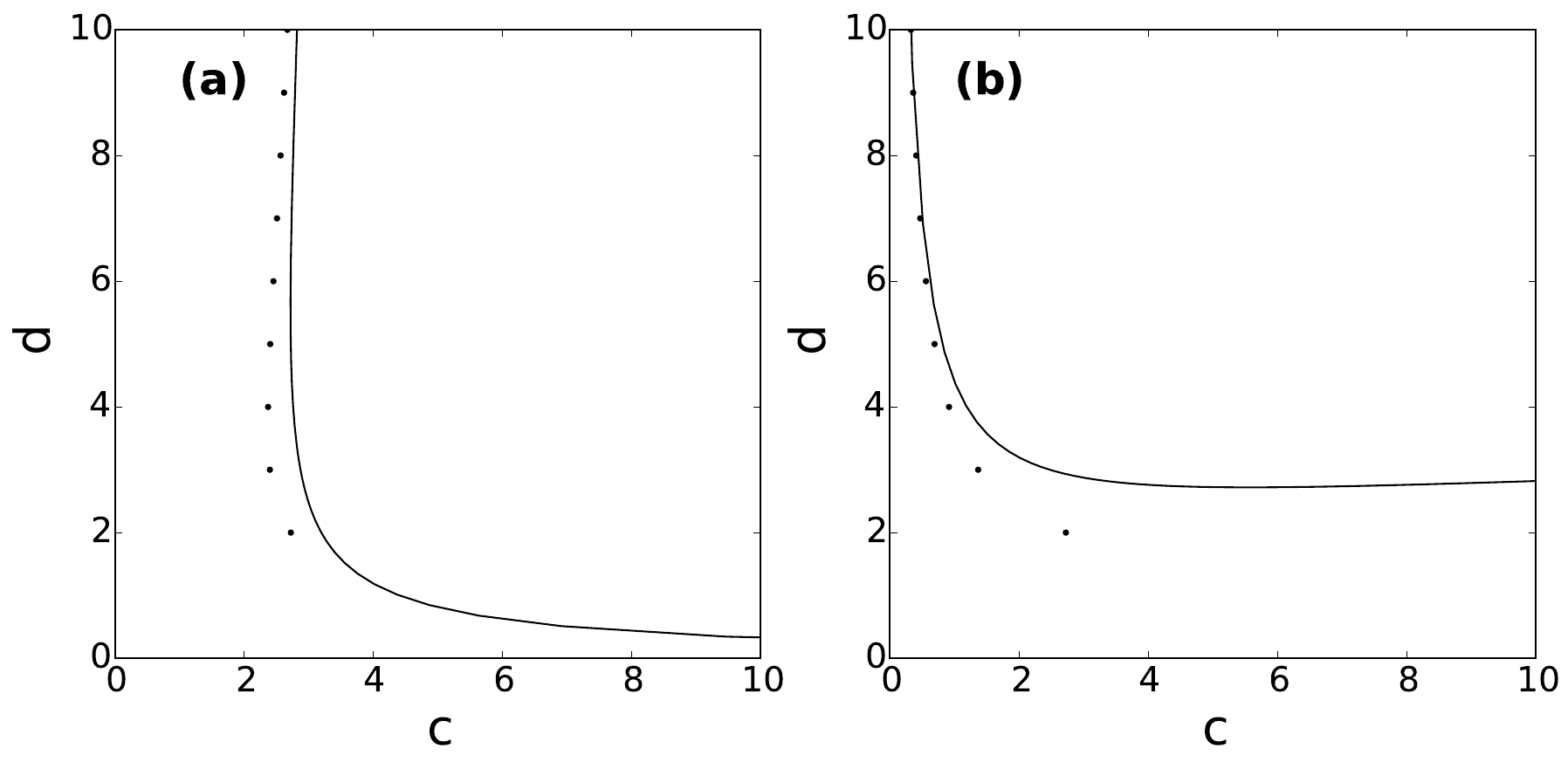}
\caption{Phase diagram of the h$_{\mm{core}}$ and v$_{\mm{core}}$
  percolations on hypergraphs with Poisson vertex degree
  distributions. Black dots and black line represenst the phase
  boundary of $d$-uniform hypergraphs and hypergraphs with Poisson
  hyperedge cardinality distribution, respectively. 
(a) h$_{\mm{core}}$. 
Note that, for $d$-uniform hypergraph ($d>1$)
with Poisson vertex degree distribution, 
the critical mean degree (i.e., $c^*$ of the v$_\mm{core}$
percolation) can be simply related to $d$ as  
$c^*=\frac{e}{d-1}$,  
where $e =2.71828\cdots$ (see Fig.~\ref{fig4}(a) black dots and SI Sec.IV.C for details). This result
was previously found in~\cite{citeulike:11962060}. 
%
(b) v$_{\mm{core}}$. 
\label{fig4}}
\end{figure}

Since the hypergraph core h$_\mm{core}$ is closely
related to the MHC problem, we study the corresponding core
percolation problem on random hypergraphs. To achieve that, we
generalize the mean-field approach proposed for the graph
case~\cite{PhysRevLett.109.205703}. We define two types of removable
vertices: a vertex is (i) $\alpha$-removable if it is or can become a
vertex of degree one; (ii) $\beta$-removable if its degree
is larger than one and belongs to at least one leaf hyperedge. 
Dually, we define two types of removable hyperedges: a hyperedge is 
(i) $\delta$-removable if it is or can become an
leaf hyperedge;
(ii) $\epsilon$-removable if it has cardinality $r$ and
is removed because it is connected to $(r-1)$ $\beta$-removable
vertices.  
Consider a large uncorrelated random hypergraph $\mathcal{H}$ with
arbitrary vertex degree and hyperedge cardinality
distributions. We can determine the
category of a vertex $v$ in $\mathcal{H}$ by the categories of its
neighboring hyperedges in the modified hypergraph $\mathcal{H} \setminus v$ with
vertex $v$ and all its hyperedges removed from $\mathcal{H}$, using the following rules:
(i) $\alpha$-removable vertex: all neighboring hyperedges are
$\epsilon$-removable; 
(ii) $\beta$-removable vertex: at least one neighboring hyperedge is
$\delta$-removable. 
Similarly, we can determine the
category of a hyperedge $e$ in $\mathcal{H}$ by the categories of its
neighboring vertices in the modified hypergraph $\mathcal{H} \setminus e$ with
hyperedge $e$ and all its vertices removed from $\mathcal{H}$, using the following rules:
(iii) $\delta$-removable hyperedge: at least one neighboring vertex is
$\alpha$-removable; 
(iv) $\epsilon$-removable hyperedge: at least one neighboring vertex is
$\beta$-removable. 
Let $\alpha$ (or $\beta$) denote the probability that
a random neighboring vertex of a random hyperedge $e$ in a hypergraph $\mathcal{H}$
is $\alpha$-removable (or $\beta$-removable) in $\mathcal{H} \setminus
e$. 
Let $\delta$ (or $\epsilon$) denote the probability that
a random neighbor of a random vertex $v$ in a hypergraph $\mathcal{H}$
is $\alpha$-removable (or $\beta$-removable) in $\mathcal{H} \setminus
v$. 
Then rules (i)-(iv) enable us to derive a set of self-consistent equations:  
%
\begin{align}
\alpha&=\sum_{k=1}^{\infty} Q_\mathrm{n}(k) \epsilon^{k-1},\label{core:eq1}\\
1-\beta&=\sum_{k=1}^{\infty} Q_\mathrm{n}(k) (1-\delta)^{k-1},\label{core:eq2}\\
1-\delta&=\sum_{r=1}^{\infty}  Q_h(\mathrm{r}) (1-\alpha)^{r-1},\label{core:eq3}\\
\epsilon&=\sum_{r=1}^{\infty}  Q_h(\mathrm{r})\beta^{r-1}.\label{core:eq4}
\end{align} 
Here $Q_\mathrm{n}(k)$ ($Q_\mathrm{h}(r)$) is the excess degree
(cardinality) distribution. 
The fraction of vertices in h$_{\rm core}$, denoted as $s^{\mathrm{v}}_{\mathrm{h}}$, is given by 
\begin{equation}
s^{\mathrm{v}}_{\mathrm{h}}=\sum_{k=2}^{\infty} P_\mathrm{n}(k) \sum_{l=2}^{k}  
\binom{k}{l} ( 1-\delta-\epsilon)^l \epsilon^{k-l}
\label{core:eq12}
\end{equation}
where $P_\mathrm{n}(k)$ ($P_\mathrm{h}(r)$) is the degree
(cardinality) distribution. (See SI Sec.IV.B for the formula of the
fraction of hypedeges in h$_{\rm core}$, denoted as
$s^{\mathrm{h}}_{\mathrm{h}}$.) 
For hypergraphs with Poisson vertex degree distribution and different
hyperedge cardinality distributions, we find that the h$_{\rm core}$ emerges as
a continuous phase transition (see Fig.~\ref{graph3}(a) and (b)),  
 \begin{equation}
 s^{\mathrm{v}}_{\mathrm{h}}\propto  (c-c^*)^{\zeta_1}\label{core:eq13},
 \end{equation}
with critical exponent $\zeta_1=1$ (see SI Sec.IV.E for details).
The relation between the critical mean degree $c^*$ (percolation
threshold) and the hyperedge mean cardinality $d$ is represented in
Fig.~\ref{fig4}. 
%

Similarly, the v$_\mm{core}$ percolation (associated with the MVC
problem) can also be analytically studied for random hypergraphs. 
In this case, the equations on
$\alpha$ and $\beta$ 
are the same as the h$_\mm{core}$ percolation case, but for hyperedges we derive the following
self-consistent equations: 
 \begin{align}
 \delta&=\sum_{r=1}^{\infty}  Q_\mathrm{h}(r) \alpha^{r-1},\label{core:eq7}\\
 1-\epsilon&=\sum_{r=1}^{\infty}  Q_\mathrm{h}(r)(1-\beta)^{r-1}\label{core:eq8}.
 \end{align} 
%
The v$_\mm{core}$ consists of those vertices connected to at least two
non-removable hyperedges. Hence, the fraction of vertices in
v$_\mm{core}$ is given by 
\begin{equation}
s^{\mathrm{v}}_{\mathrm{v}}=\sum_{k=2}^{\infty} P_\mathrm{n}(k) \sum_{l=2}^{k} 
\binom{k}{l} ( 1-\delta-\epsilon)^l \epsilon^{k-l}. 
\label{core:eq9}
\end{equation}
See SI Sec.IV.B for the formula
of the fraction of hypedeges in $v_{\rm core}$,
$s^{\mathrm{h}}_{\mathrm{h}}$. 
For hypergraphs with Poisson vertex degree distribution and different
hyperedge cardinality distributions, we find that the $v_{\rm core}$ emerges as
a continuous phase transition (see Fig.~\ref{graph3}(c,d)): 
\begin{equation}
 s^{\mathrm{v}}_{\mathrm{v}} \sim (c-c^*)^{\zeta_2}\label{core:eq11}, 
\end{equation}
%
with critical exponent $\zeta_2=1$ (see SI Sec.IV.D for details).
Fig.~\ref{graph3}~(d) shows that for a Poisson-Poisson hypergraph the
size of $s^{\mathrm{v}}_{\rm v}$ starts to decrease at large values of
$c$. By increasing the number of hyperedges connected to a vertex, but
keeping the cardinailty distribution constant, the probability of a
vertex being connected to a hyperedge with cardinality one increases,
and any vertex connected to a  hyperedge with cardinality one is
automatically removed.  This effect is not relevant if the probability
that a vertex is connected to a  hyperedge with cardinality one is
very small,  $1-\exp(-c~e^{-d})\ll1$. For large values of $c$ and $d$,
this effect is only relevant if $c\sim e^d$.

Phase diagrams of the h$_{\mm{core}}$ and v$_{\mm{core}}$ percolations
on hypergraphs with Poisson vertex degree distributions are shown in
Fig.~\ref{fig4}. 
Note that the phase diagram of v$_\mm{core}$ percolation is equal to that of
h$_\mm{core}$ percolation if we interchanged the mean cardinality $d$ with the
mean degree $c$. This is true because the v$_\mm{core}$ of a
hypergraph is the h$_\mm{core}$ of the dual hypergraph.  
%

%
%

\begin{figure}[t!]
\centering
\includegraphics[width=0.45 \textwidth]{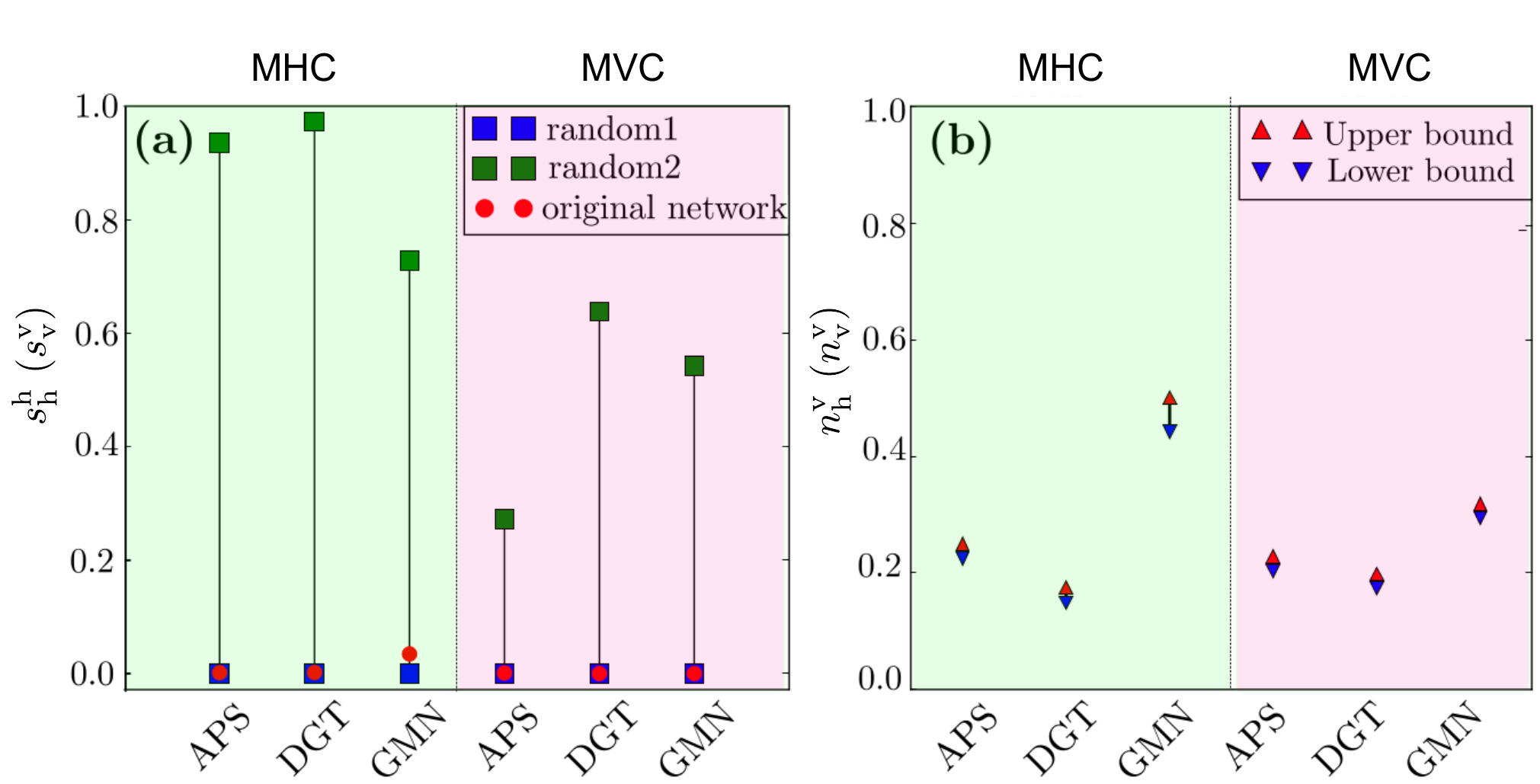}
\caption{
\label{real_data_fig2} (a) Fraction of hyperedges $s^{\mathrm{h}}_{\rm
  h}$ (or vertices $s^{\mathrm{v}}_{\rm v}$) associated with the MHC
(MVC) problem for three real-world hypergraphs: APS, DGT and GMN; and
their respective randomized counterparts. This set represents those
hyperedges (vertices) that cannot be covered optimally  
using our generalized GLR procedure. (b) Fraction of vertices
$n^{\mathrm{v}}_{\rm h}$ (hyperedges $n^{\mathrm{h}}_{\rm v}$) necessary to
cover all the hyperedges (vertices) for the three hypergraphs.}
\end{figure}
%
We also apply the generalized GLR procedure to compute the h$_{\mm{core}}$
and v$_{\mm{core}}$ for several real-world hypergraphs: 
(1) APS consists of articles published in all the Americal
Physical Society journals from 1893 to
2010~\cite{sinatra2014scientific}, where individual authors and their
joint articles are considered as vertices and hyperedges,
respectively. 
(2) DGT consists of drugs (hyperedges) and their target genes
(vertices) as listed in the DrugBank~\cite{DrugBank}. 
(3) GMN consists of reactions (hyperedges) and the involved
metabolites (vertices) in the genome-scale metabolic network of
{\it{E. coli}} obtained from the BiGG
database~\cite{Schellenberger2010}. 
We find that for GMN, h$_\mm{core}$ contains 3.4\% of hypereges; while
v$_\mm{core}$ contains less than 0.2\% of vertices. 
%
For the other two hypergraphs (APS and DGT), both h$_\mm{core}$ and
v$_\mm{core}$ contains less than 0.2\% of hyperedges or vertices
(Fig.~\ref{real_data_fig2}a). 
We also compare the size of each core with that of two randomized
counterparts of the real-world hypergraphs. For the first randomization
scheme (random1), we preserve
both the vertex degree and hyperedge cardinality distributions of the
real hypergraph. 
For the second randomization scheme (random2), we consider a random Poisson-Poisson hypergraph with the
same average degree and average cardinality as the real hypergraph. 
Note that for random1, the size of the core is always zero (blue
points in Fig.~\ref{real_data_fig2}a); while for random2, the size of the core is between 30\% and
100\%, much bigger than that of the real hypergraph. 
These results suggest that the degree and cardinality distributions
are main factors that explains the small cores of these real-world
hypergraphs.  
Because h$_\mm{core}$ and v$_\mm{core}$ of those real-world hypergraphs
are really small, the MHC and MVC problems are effectively
solvable. Indeed, as shown in Fig.~\ref{real_data_fig2}(b), the upper and lower
bound of the fraction of vertices (hyperedges) $n^{\mathrm{v}}_{\rm
  h}$ ($n^{\mathrm{v}}_{\rm v}$) that are necessary to cover all the
vertices (hyperedges) are very close to each other.

\begin{figure}[t!]
\centering
\includegraphics[width=0.45\textwidth]{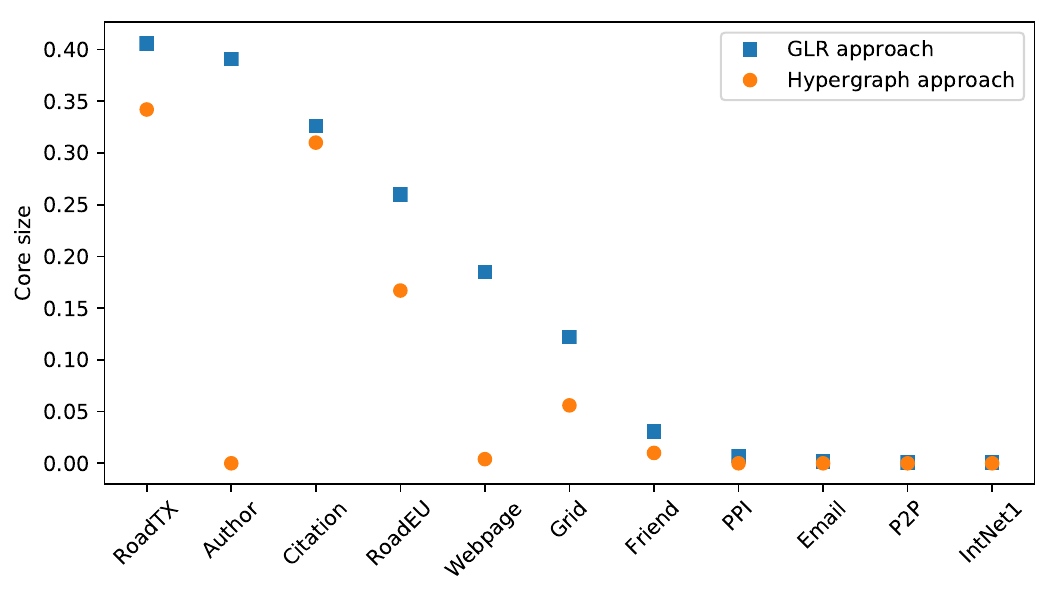}
\caption{Relative size of the core associated with the MDS problem for
  eleven real-world networks, using the GLR procedure introduced in
  \cite{Zhao2015} (square) and using our generalized GLR approach
  (circles). \label{table1}}  
\end{figure}

It turns out that our method can also be used to solve another
classical NP-hard combinatorial problem: finding the minimum dominating set
(MDS) for graphs. 
The MDS of a graph is the smallest set of vertices that needs to be
occupied so that all unoccupied vertices are adjacent to at least one
occupied vertex~\cite{trove.nla.gov.au/work/24617048}. 
Our basic idea is as follows.  We consider a hypergraph that has the
same set of vertices as the original graph, and one hyperedge $i$
contains all vertices adjacent to a vertex $v_i$ (including  $v_i$ 
itself). Solving the MHC problem on this hypergraph is
then equivalent to solve the MDS problem of the original
graph. 
Our method offers a much more general approach than the greedy algorithm
introduced in~\cite{Zhao2015}. Indeed, the  two leaf-removal rules
introduced in ~\cite{Zhao2015} can be considered as special cases of
our generalized GLR rules (see SI Sec.IV.A for details). In
Fig.~\ref{table1}, we show the size of the cores
associated with the dominating set for the eleven real-world networks analyzed
in \cite{Zhao2015}. For most of those networks, our method shows a considerable
improvement over the previous method. For some of the networks, our
method actually finds no core left.  

%
%
%
%
In physics it is common that a more abstract or general approach
actually makes certain complicated problems easier to solve. This is
often not true in social systems, biological systems, or complex
systems in general. Our results suggest that generalizing graph to 
hypergraphs is one of the few cases where a small generalization makes
a very hard problem easier to solve. Indeed, we show that our
generalized GLR procedure and the corresponding hypergraph cores can
help us solve various NP-hard covering problems in a systematic and
universal way. If we aim to find a simple solution for complex problems, this
is really an exciting result, indicating that hypergraphs might be the
right way to represent complex networked systems. 
Our results open a new set of tools to analyze complex networked
systems. It also raises a very important question: why do hypergraph
cores of real systems tend to be small or absent? We anticipate our
work will trigger more research activities in addressing this
intriguing question.

\begin{acknowledgments}
This work was partially supported by the
Funda{\c c}{\~ a}o para a Ci{\^ e}ncia e a Tecnologia (Portugal)
through grant No. SFRH/BD/79723/2011 and through projects
UID/EEA/50008/2013 and UID/EEA/50008/2019. 
\end{acknowledgments}

\end{document}


\title{Supplementary Information}
\author{Bruno Coelho Coutinho}
\affiliation{Instituto de Telecomunica\c{c}\~{o}es, Physics of Information and Quantum Technologies Group, Portugal}

\author{Hai-Jun Zhou}
\affiliation{CAS Key Laboratory for Theoretical Physics, Institute of Theoretical Physics, Chinese Academy of Sciences, Beijing 100190, China} 
  \author{Yang-Yu Liu}
\affiliation{Channing Division of Network Medicine, Brigham and
  Women's Hospital and Harvard Medical School, Boston, Massachusetts 02115, USA} 
\affiliation{Center for Cancer Systems Biology, Dana-Farber Cancer
  Institute, Boston, Massachusetts 02115, USA}

\maketitle
\tableofcontents
\clearpage


\clearpage

\section*{Organization}
In Sec.\ref{section1} we introduce the concept of a hypergraph and different types of cores and percolation processes in graphs. In Sec.\ref{section2} we explore the giant connected component and in Sec.\ref{section3} a generelized version of K-core percolation for hypergraphs. Although not fundamental to the main paper, it provides standard tools to  study percolation transitions in hypergraphs. 
In Sec.\ref{section4} he show detailed calculation underlying the results in the main text. 
In Sec.\ref{section5} we compare the results of our analytical calculation with the results of numerical simulations.

\clearpage
\section{Introduction}
\label{section1}



\begin{figure}[!t]
\centering
\includegraphics[width=1\textwidth]{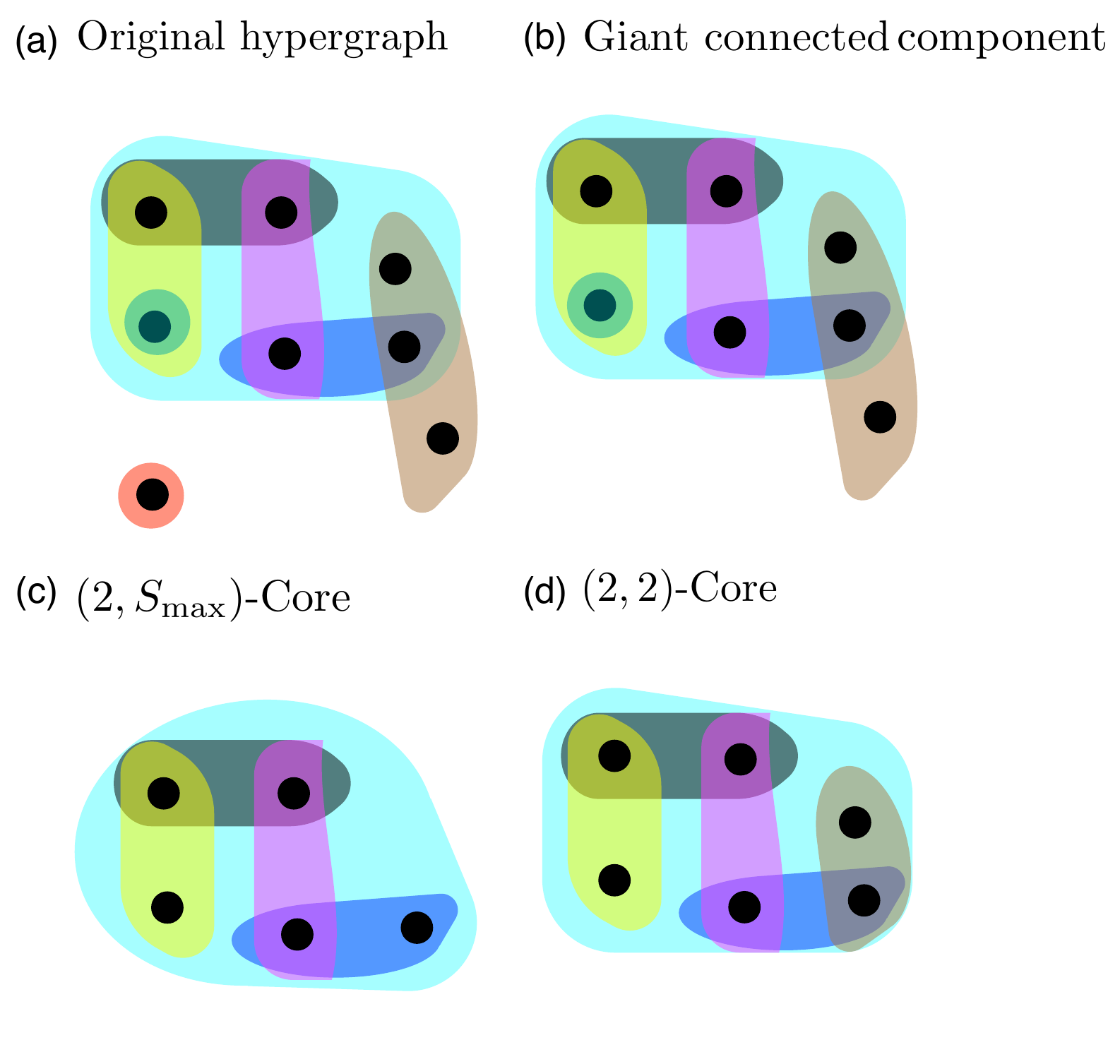}
\caption{ (Color online) Example of the different percolations studied in the work. (a) shows the original hypergraph. (b), (c) and (d) shows the $(2,S_{\max})$-core, $(2,2)$-core, where  $S_{\max}\equiv \max(r,2)$ .
}  \label{toy}
\end{figure}

Despite the ubiquity of hypergraphs in different fields,
fundamental structural properties of hypergraphs have not been fully
understood. Most of the previous works focus on uniform
hypergraphs~\cite{Goldschmidt05criticalrandom,Cooper03thecores,citeulike:11962060}, ignoring the fact that hyperedges could have a wide range
of cardinalities. 
In this work, we systematically study the percolation transitions on
hypergraphs with arbitrary vertex degree and hyperedge cardinality
distributions.  We are particularly interested in the 
emergence of a giant component, the $K$-core, and the core in
hypergraphs (see Fig.~\ref{toy}).   
Those special subgraphs have been extensively studied
in the graph case and play very important roles in many network
properties \cite{citeulike:1263642,Alvarez-hamelin06largescale}. 
A giant component of a graph is a connected component that contains a
finite fraction of the entire graph's vertices, which is relevant to
structural robustness and resilience of networks \cite{RevModPhys.80.1275,citeulike:4012374}.   
The $K$-core of a graph is obtained by recursively removing vertices
with degree less than $K$, as well as edges incident to them. The
$K$-core has been used to identify influential spreaders in complex
networks \cite{Gallos}.  
The core of a graph is the remainder of the greedy leaf
removal (GLR) procedure: leaves (vertices of degree one) and their
neighbors are removed iteratively from the graph. The emergence of the
core in a graph has been related to the conductor-insulator transition
\cite{PhysRevLett.86.2621}, structural controllability~\cite{liu11}, and many combinatorial optimization problems \cite{4568355}.




We can naturally extend the definition of giant component to the
hypergraph case. Yet, to obtain the $K$-core in a hypergraph, we have
to specify how to remove hyperedges containing vertices of degree less
than $K$.   
To achieve that, we introduce the $(K,S)$-core defined as the largest fraction of the hypergraph where each hyperedge contains at least $S$ vertices and each vertex belongs to at least $K$ hyperedges in the subset. The $(K,S)$-core is obtained by recursively removing vertices with degree less than $K$ and hyperedges with cardinality less than $S$.

\clearpage

\section{Giant component} 
\label{section2}
A giant connected component of a hypergraph is a connected component that
contains a constant fraction of the entire vertices. 
In the mean-field picture, we can derive a set of self-consistent
equations to calculate the relative size of the GCC, using the
generating function formalism \cite{citeulike:48}. 
%
%
%
Let $\mu$ represent the probability that a randomly selected vertex from a
randomly chosen hyperedge is not connected via other hyperedges with
the GCC. 
Dually, let $\psi$ represent the probability that a randomly chosen hyperedge
connecting to a randomly chosen vertex is not connected via other vertices
with the GCC. 
Then we have 
\begin{align}
&\mu=\sum_{k=1}^{\infty} Q_\mathrm{n}(k) \psi^{k-1}\label{per:eq1_1}\\
&\psi=\sum_{r=1}^{\infty} Q_\mathrm{h}(r) \mu^{r-1}.\label{per:eq1_2}
\end{align}
%
Here $Q_\mathrm{n}(k)\equiv k  P_\mathrm{n}(k) /c$ is the excess
degree distribution of vertices, i.e., the degree distribution for the
vertices in a randomly chosen hyperedge. 
$P_\mathrm{n}(k)$ is the vertex degree distribution, and $c=c_1$ is the
mean degree of the vertices. In general we define  $c_m \equiv
\sum_{k=0}^{\infty} k^m  P_\mathrm{n}(k)$.  
$Q_\mathrm{h}(r)\equiv r P_{\rm h}(r) /d$ is the excess cardinality
distribution of hyperedges, i.e., the cardinality distribution for
the hyperedges connected to a randomly chosen vertex. 
$P_{\rm h}(r)$ is the hyperedge cardinality distribution, and $d=d_1$ is the
mean cardinality of the hyperdges. In general we define  $d_m  \equiv
\sum_{r=0}^{\infty} r^m  P_{\rm h}(r)$.

\begin{figure}[!htb]
\centering
\includegraphics[width=1\textwidth]{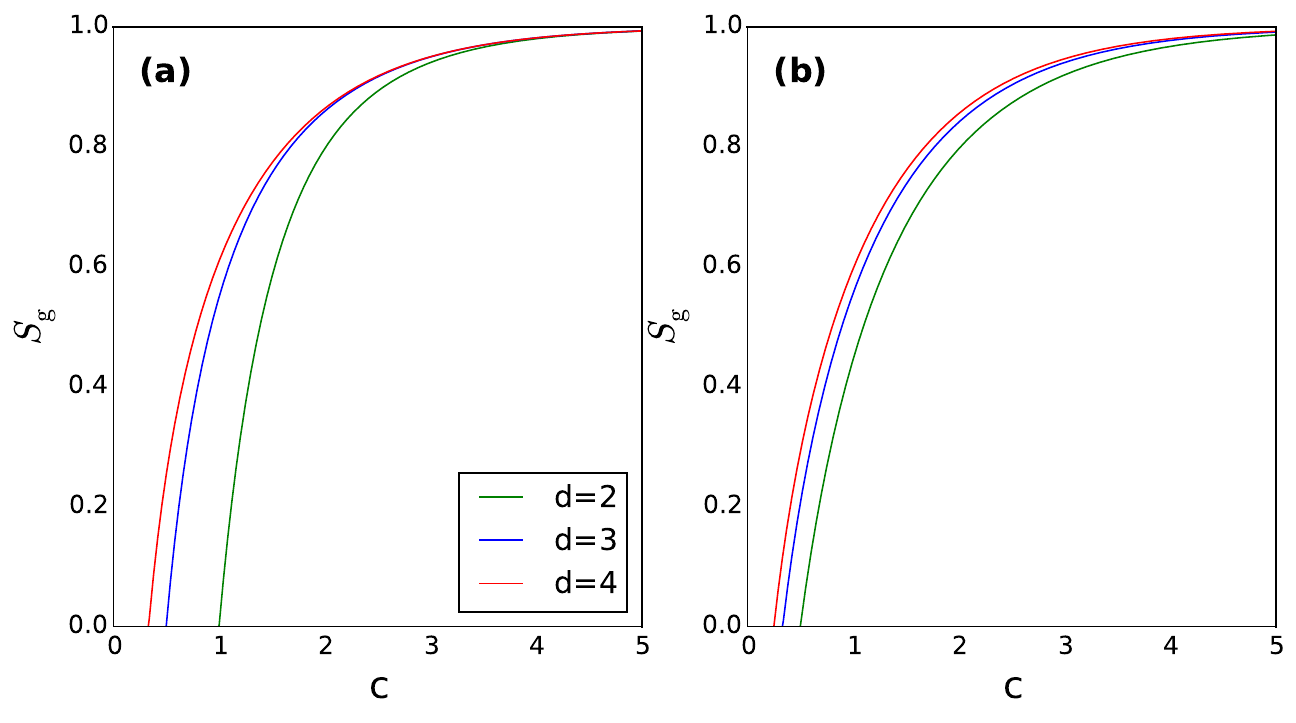}
\caption{ (Color online) The relative size of the GCC
  $s_\mathrm{g}$ as a function of the mean degree $c$ for hypergraphs
  with Poisson vertex degree distribution. (a) $d$-uniform hypergraphs with $d=2,3,4$; (b)
  hypergraphs with Poisson hyperedge cardinality distribution and mean
  cardinality $d=2,3,4$.\label{graph1b}
}
\end{figure}

The relative size of the GCC is then given by 
\begin{equation}
s_\mathrm{g}=1-\sum_{k=0}^{\infty}P_\mathrm{n} (k) \psi^{k} \label{per:eq3}.
\end{equation}
Fig.~\ref{graph1b} shows the analytical result of $s_\mathrm{g}$ as a
function of the mean degree $c$ for hypergraphs with Poisson vertex
degree distribution and different hyperedge cardinality distributions.  
%
%
%
Clearly the giant component in hypergraphs emerges as a continuous 
phase transition with scaling behavior
\begin{equation}
s_\mathrm{g} \sim  (c-c^*)^\eta\label{per:eq5}
\end{equation}
for $c-c^*\rightarrow 0^{+}$,  where $c^*$ is the critical value of
mean degree (i.e., the percolation threshold) and $\eta$ is the
critical exponent associated with the critical singularity.


The condition for the percolation transition can be determined by rewriting Eqs. ~\ref{per:eq1_1} as
\begin{equation}
F_\mathrm{g}=0,
\label{eq_fg}
\end{equation}
where we define
\begin{equation}
F_\mathrm{g}\equiv\sum_{k=1}^{\infty} Q_\mathrm{n}(k) \psi^{k-1}-\mu.
\label{eq_fg}
\end{equation} 
At the critical point when $\mu^*=1$ and
\begin{equation}
\left.\partiald{F_\mathrm{g}}{\mu}\right|_{\mu=\mu^*=1}=0,
\label{eq1}
\end{equation}
we obtain
\begin{equation}
\frac{ d_2-d}{d}\frac{c_2 - c}{c}>1.\label{per:eq4}
\end{equation}
where $c^*$ and $c^*_2$ are the hyperdegree first and second moment, respectively, at the critical point.

Note that a similar relation has been found for uniform
hypergraphs~\cite{Mezard2009IPC1592967}. In the graph case ($d=2$ for
all edges) we recover the classical result $\frac{c_2 -
  c}{c}>1$ \cite{RevModPhys.80.1275,molloyreed95}.

The critical exponent $\eta$ can be calculated by consider a point around the critical point, such as, $\mu=1-\zeta$ with $\zeta=0^+$  and  $c=c^*+\chi$ with $\chi=0^+$. We can  define the function from Eq. ~\ref{eq_fg} as a function of  $\mu$ and $c$  (i.e. $F_\mathrm{g}(\mu,c)$). By expanding $F_\mathrm{g}(\mu,c)$ around the point $(\mu,c)=(1,c^*)$ and combining it with the result from Eq. ~\ref{eq1} at the critical point, we can rewrite  Eq.~\ref{eq_fg} as
\begin{eqnarray}
&&\left.\frac{\partial F_\mathrm{g}(\mu,c)}{\partial c}\right|_{(1,c^*)} \chi+\left.\frac{\partial^2 F_\mathrm{g}(\mu,c)}{\partial^2 \mu}\right|_{(1,c^*)} \zeta^2+\left.\frac{\partial^2 F_\mathrm{g}(\mu,c)}{\partial^2 c}\right|_{(1,c^*)} \chi^2\\\nonumber
&&-2\left.\frac{\partial^2 F_\mathrm{g}(\mu,c)}{\partial \mu\partial c}\right|_{(1,c^*)} \chi\zeta+(\cdot \cdot \cdot)=0.
\label{Twocore_t:eq7}
\end{eqnarray}
For $\mu=1$, 
\begin{equation}
 F_\mathrm{g}(1,c) =1-c/c=0,
\label{eq3_b}
\end{equation}
it follows that
\begin{equation}
\left.\frac{\partial^n F_\mathrm{g}(\mu,c)}{\partial^n c}\right|_{(1,c^*)} =0
\label{eq3}
\end{equation}
for any positive integer $n$.
%
Let us assume that there are no diverging moments for the hyperedge cardinality distribution. We can truncate our expansion of $ F_\mathrm{g}$ at order 2 for $\zeta$ and $\chi$. This implies that
\begin{equation}
\zeta\sim\chi.
\label{eq4}
\end{equation}
Let us expand $s_g$ as a function of $\zeta$,
\begin{equation}
s_g=\left. s_g(\mu)\right|_{\mu=\mu^*}-\left.\partiald{s_g(\mu)}{\mu}\right|_{\mu=\mu^*}\zeta+\left.\partialdt{s_g(\mu)}{\mu}\right|_{\mu=\mu^*}\zeta^2+(\cdot\cdot\cdot).
\label{eq4}
\end{equation}
For $\mu^*=1$, $s_g(\mu^*)=0$. We obtain 
\begin{equation}
s_{_\mathrm{g}}\sim  \chi,
\label{eq5}
\end{equation}
This correspond to he same exponent $\eta=1$ as in the graph case~\cite{RevModPhys.80.1275,Bollobás2013}.

\clearpage
\section{$(K,S)$-core on hypergraphs}
\label{section3}
%
%
The $(K,S)$-core of a hypergraph is obtained by recursively removing vertices
with degree less than $K$ and hyperedge with cardinality less than $S$. 
A hyperedge with cardinality $r$ is removable if at least $r-S+1$ vertices connected to it are also
 removable and a vertex with degree $k$ is removable if at least $k-K+1$ hyperedges connected to it are also
 removable. One can remove a vertex or a hyperedge from the hypergraph,
 and see what is the probability of a neighboring hyperedge or vertex,
 respectively, being removable. This allows us to derive a set of
 self-consistent equations:
 \begin{align}
&\alpha=\sum_{k=1}^{\infty} Q_\mathrm{n}(k) \sum_{l=k+1-S}^{k-1}  \left( \begin{array}{c} k-1 \\ l\end{array}\right)\delta^{l}(1-\delta)^{k-1-l},\label{2-core:eq1_0}\\
&\delta=\sum_{r=1}^{\infty} Q_\mathrm{h}(r) \sum_{l=r+1-K}^{r-1} \left( \begin{array}{c}  r-1 \\l\end{array}\right) \alpha^{l}(1-\alpha)^{r-1-l}.\label{2-core:eq1_2}
\end{align}
Here $\alpha$ and $\delta$ are, respectively, the probability that a
vertex or a hyperedge is removable. 
From now on we will focus on the case of
  $K=2$. Then Eq.~(\ref{2-core:eq1_0}) reduces to 
%
  \begin{align}
&\alpha=\sum_{k=1}^{\infty} Q_\mathrm{n}(k) \delta^{k-1}.\label{2-core:eq1_1}
\end{align}


%
\subsection{$K=2$ and $S=S_{\max}$}
\begin{figure}[t!]
\centering
\includegraphics[width=1\textwidth]{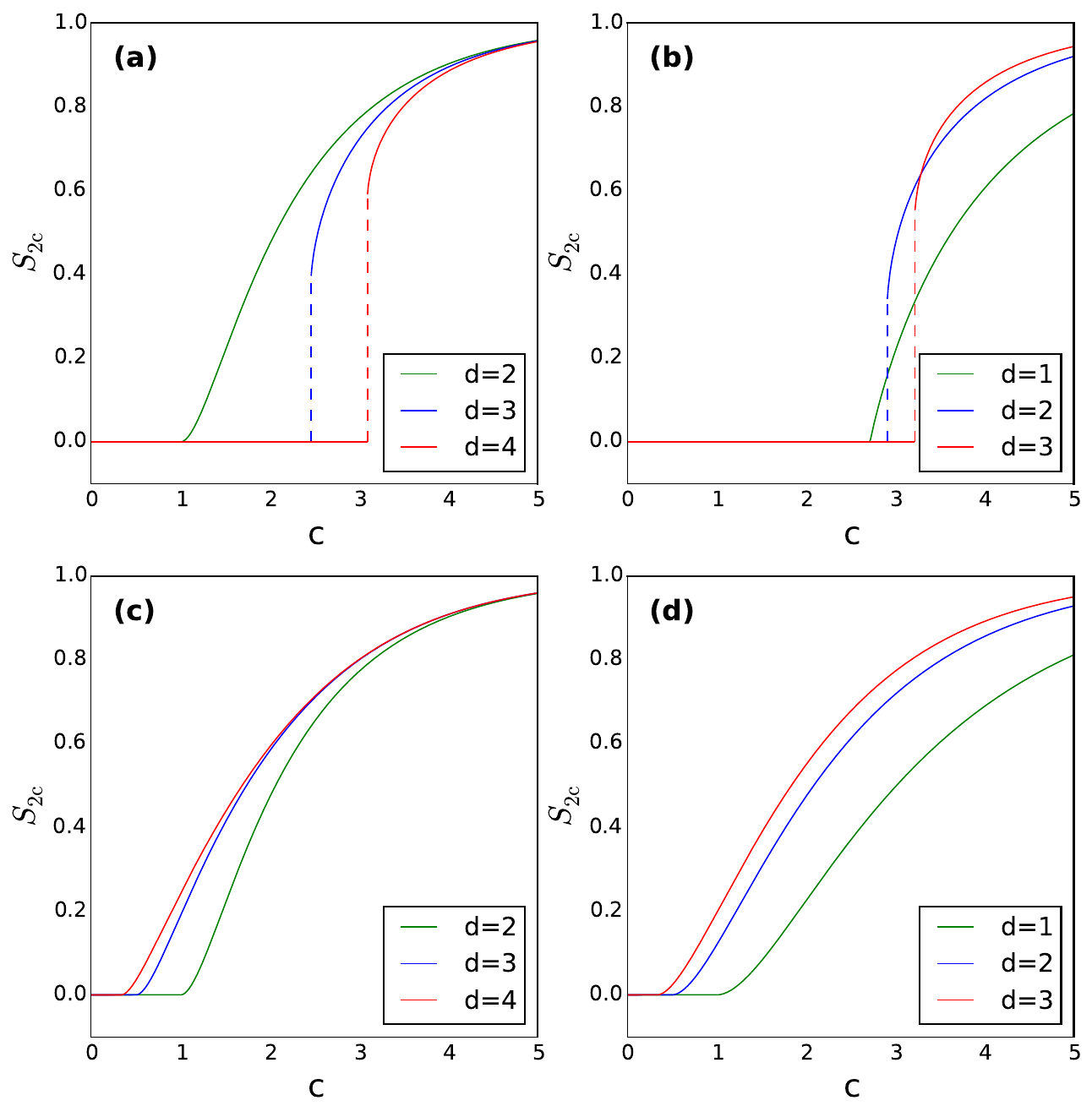}
\caption{(Color online) The relative size of $(2,S)$-core $s_{\mathrm{2c}}$ as a
  function of the mean degree $c$ for hypergraphs with Poisson degree
  distribution, for $S=\max(r-1,2)$  (a) and (b), and $S=2$, (b) and (d). 
  (a) and (c) $d$-uniform hypergraphs . (b) and (d) hypergraphs
  with Poisson hyperedge cardinality distribution.
\label{graph1c} } 
\end{figure}

The $(K,S)$-core defined with $K=2$ and $S=S_{\max}\equiv \max(r,2)$, (where $S_{\max}=2$ if $r<2$ and and $S_{\max}=r$ if $r\geq  2$ and $r$ is the intial cardinality of an hyperedge before any removal process) is
obtained by recursively removing all vertices with degree one as well as the
hyperedges containing them, and all hyperedges with cardinality smaller than two. Hyperedges with cardinality one or zero do not connect any nodes, thus have no meaning in what cores are concerned.  In this case the threshold $S$ depends on the cardinally of the hyperedges. Furthermore, if one of the  vertices of any hyperedge is removed the hyperedge is also removed. 
(Note that the $(2,S_{\max})$-core 
has been defined in literature~\cite{Cooper03thecores} simply as $2$-core,
and discontinuous $2$-core percolation is found in $d$-uniform
hypergraphs with $d>2$.)  
In this case, Eq.~(\ref{2-core:eq1_2}) reduces to 
\begin{equation}
1-\delta=\sum_{r=2}^{\infty} Q_\mathrm{h}(r) (1-\alpha)^{r-1}.\label{2-core:eq1_3}
\end{equation}
The relative size of the $(2,S_\mm{max})$-core is given by the probability that a
randomly chosen vertex is connected to at least two non-removable
hyperedges: 
\begin{equation}
s_\mathrm{2c}=\sum_{k=2}^{\infty} P_{\rm h}(k) \sum_{l=2}^{k} \left( \begin{array}{c} k \\ l\end{array}\right)( 1-\delta)^l \delta^{k-l}.
\label{2-core:eq1_5}
\end{equation}
Fig.~\ref{graph1c} shows the analytical result of $s_\mathrm{2c}$ as a
function of the mean degree $c$ for hypergraphs with Poisson vertex
degree distribution and different hyperedge cardinality distributions.  
We find that, depending on the mean hyperedge cardinality $d$, the
$(2,S_{\max})$-core emerges as either a continuous or a hybrid phase transition,
with scaling behavior 
\begin{equation}
s_\mathrm{2c} -s_\mathrm{2c}^*\sim  (c-c^*)^\zeta\label{2-core:eq1_4}
\end{equation}
for $c- c^* \to 0^+$, where $c^*$ is the percolation
threshold and $\zeta$ is the critical exponent. 
$s_\mathrm{2c}^*$ is the $(2, S_\mm{max})$-core relative size right at the
critical point: $s_\mathrm{2c}^*=0$ for continuous phase transition
and non-zero for hybrid phase transitions. 
%
%
The percolation threshold $c^*$ can be calculated by defining
\begin{equation}
F_\mathrm{2c}\equiv\sum_{k=1}^{\infty} Q_\mathrm{n}(k) \delta^{k-1}-\alpha=0.
\label{eq_f2c}
\end{equation}
If we consider $S=\max(r,2)$, $\delta$, defined in Eq.~\ref{2-core:eq1_3}, and we  combine Eqs.~\ref{eq_f2c}, ~\ref{eq6_chapter5} and \ref{2-core:eq1_3}, we obtain
\begin{equation}
1=\sum_{k=1}^{\infty} \sum_{r=2}^{\infty} Q^*_\mathrm{n}(k)Q_\mathrm{h}(r) (k-1)(r-1){\delta^*}^{k-2}(1-\alpha^*)^{r-2},\label{eq7}
\end{equation} 
where $Q^*_\mathrm{n}(k)$ is the connectivity distribution with the critical mean hyperdegree  $c^*$. $\delta^*$ and  $\alpha^*$ are the values of $\delta$ and $\alpha$ at the critical point, respectively. The phase transition is continuous if for  $\delta^*=\alpha^*=1$,
\begin{equation}
\left.\partialdt{F_\mathrm{2c}}{\alpha}\right|_{\alpha=1}>0,
\label{eq8}
\end{equation}
that reduces to,
\begin{equation}
2Q_\mathrm{n}(3) (c^*_2-c^*)-Q_\mathrm{n}(2)^2 (c^*_3-3c^*_2+2c^*)<0.
\label{2-core:eq1_11_0}
\end{equation}
where $c^*_3$ and $c^*_2$ is third and second moment of the degree distribution at the critical point, respectively. The  phase transition is only continuous if
at the critical point
\begin{equation}
\left.\partiald{F_\mathrm{2c}}{\alpha}\right|_{(\alpha,c_1)=(\alpha^*,c^*)}=0.
\label{eq6_chapter5}
\end{equation}
From this condition we obtain that the critical point for a continuous phase transition is given by,
\begin{equation}
Q_\mathrm{h}(2) \frac{c^*_2- c^*}{c^*}=1.
\label{2-core:eq1_10_0}
\end{equation} 
Let us assume a Poisson distribution of hyperdegrees, in this case $c^*_2-c^*={c^*}^2$ and $c^*_3-3c^*_2+2c^*={c^*}^2$. Combine these relations with Eqs.~\ref{2-core:eq1_11_0} and \ref{2-core:eq1_10_0}, Eq. \ref{2-core:eq1_10_0} reduces to
\begin{equation}
2Q_\mathrm{h}(3)<Q_\mathrm{h}(2).
\label{2-core:eq2}
\end{equation}
If the hyperedge cardinality follows a Poisson distribution, there is a continuous phase transition when $d_1<1$,  witch corresponds to
\begin{equation}
d_1<\bar{d},
\label{condition}
\end{equation}
where $\bar{d}\equiv 1$.
Let us consider points around the critical point, such as $\alpha=\alpha^*-\zeta$ with $\zeta=0^+$  and  $c=c^*+\chi$ with $\chi=0^+$. We can  define the function from Eq.~\ref{eq_f2c} as a function of  $\alpha$ and $\chi$ (i.e. $F_\mathrm{2c}(\alpha,c)$). By expanding $F_\mathrm{g}(\alpha,\chi)$ around the point $(\alpha,c)=(\alpha^*,c^*)$ and combining it with result from Eq~\ref{eq6_chapter5} at the critical point, we can rewrite  Eq.~\ref{eq_fg} as
\begin{equation}
\left.\frac{\partial F_\mathrm{2c}(\alpha,c)}{\partial c}\right|_{(\alpha^*,c^*)} \chi+\left.\frac{\partial^2 F_\mathrm{2c}(\alpha,c)}{\partial^2 \alpha}\right|_{(\alpha^*,c^*)} \zeta^2+\left.\frac{\partial^2 F_\mathrm{2c}(\alpha,c_1)}{\partial^2 c}\right|_{(\alpha^*,c^*)} \chi^2-2\left.\frac{\partial^2 F_\mathrm{2c}(\alpha,c)}{\partial \alpha\partial c}\right|_{(\alpha^*,c^*)} \chi\zeta+(\cdot \cdot \cdot)=0.
\label{giant:eq7}
\end{equation}
For the continuous phase transition, $\alpha^*=1$, this implies
\begin{equation}
\left.\frac{\partial^n F_\mathrm{2c}(\mu,c_1)}{\partial^n c}\right|_{(1,c^*)} =0
\label{chapter5_eq9}
\end{equation}
for any positive integer $n$. Thus,
\begin{equation}
\zeta\sim  \chi.
\label{eq11}
\end{equation}
Let us expand $s_{2c}$ as a function of $\zeta$,
\begin{equation}
s_{2c}=\left. s_{2c}(\mu)\right|_{\alpha=\alpha^*}-\left.\partiald{s_{2c}(\alpha)}{\alpha}\right|_{\alpha=\alpha^*}\zeta+\left.\partialdt{s_{2c}(\alpha)}{\alpha}\right|_{\alpha=\alpha^*}\zeta^2+(\cdot\cdot\cdot),
\label{eq4}
\end{equation}
For $\alpha^*=1$, 
$s_{2c}(\alpha^*)=0$ and
\begin{equation}
\left.\partiald{s_{2c}(\alpha)}{\alpha}\right|_{\alpha=\alpha^*=1}=0.\label{2c_eq1}
\end{equation}
We obtain
\begin{equation}
s_{2c}\sim  \zeta^2 \Rightarrow s_{2c}\sim  \chi^2,
\label{eq12}
\end{equation}
for $s_{2c}^*=0$ and $\eta=2$.
For the discontinuous phase transition, $\alpha^*\neq1$, we have
\begin{equation}
\left.\frac{\partial F_\mathrm{2c}(\alpha,c)}{\partial c}\right|_{(1,c^*)} \neq0,
\label{eq13}
\end{equation}
$s_{2c}(\alpha)\neq 0$ and
\begin{equation}
\left.\partiald{s_{2c}(\alpha)}{\alpha}\right|_{\alpha=\alpha^*}\neq0.
\end{equation}
It implies
\begin{equation}
\zeta\sim  \chi^{1/2},
\label{eq14}
\end{equation}
and
\begin{equation}
s_{2c}-s^*_{2c}\sim  \chi^{1/2},
\label{eq15}
\end{equation}
that is equivalent to Eq.~\ref{2-core:eq1_4} for $\eta=1/2$. A discontinuous phase transition with a critical exponent smaller than one is considered as hybrid phase transition.\\

At the critical  mean cardinality, $d=\bar{d}$, $\alpha^*=1$,  but
\begin{equation}
\left.\partialdt{F_\mathrm{2c}}{\alpha}\right|_{\alpha=1}=0.
\label{eq8_chapter5}
\end{equation}
In this case we have to take into account terms of order $3$ in the expansion of Eq.~\ref{giant:eq7}. We obtain, 
\begin{equation}
\zeta\sim  \chi^{1/2},
\label{eq14}
\end{equation}
and,
\begin{equation}
s_{2c}\sim  \chi,
\label{eq12}
\end{equation}

 We can summarize that  for $d$-uniform hypergraphs the $(2,S_\mm{max})$-core percolation is
(i) continuous with critical exponent $\zeta=2$ if $d=2$; and (ii) hybrid with
critical exponent $\zeta=1/2$ if $d>2$ (which is consistent with a previous
work~\cite{Cooper03thecores}). 
For hypergraphs where both the vertex degeree and hyperedge cardnality
distributions are Poissonian, the $(2,\max(r,2))$-core percolation is (i) continuous
with critical exponent $\zeta=2$ if $d<\bar{d}=1$; (ii)
continous with critical exponent $\zeta=1$ if $d=\bar{d}$; and  (iii)
hybrid with critical exponent $\zeta=1/2$ if $d>\bar{d}$. The same set of critical exponents was found for the heterogeneous-$K$-core \cite{PhysRevLett.107.175703}.
\begin{figure}[t!]
\centering
\includegraphics[width=1\textwidth]{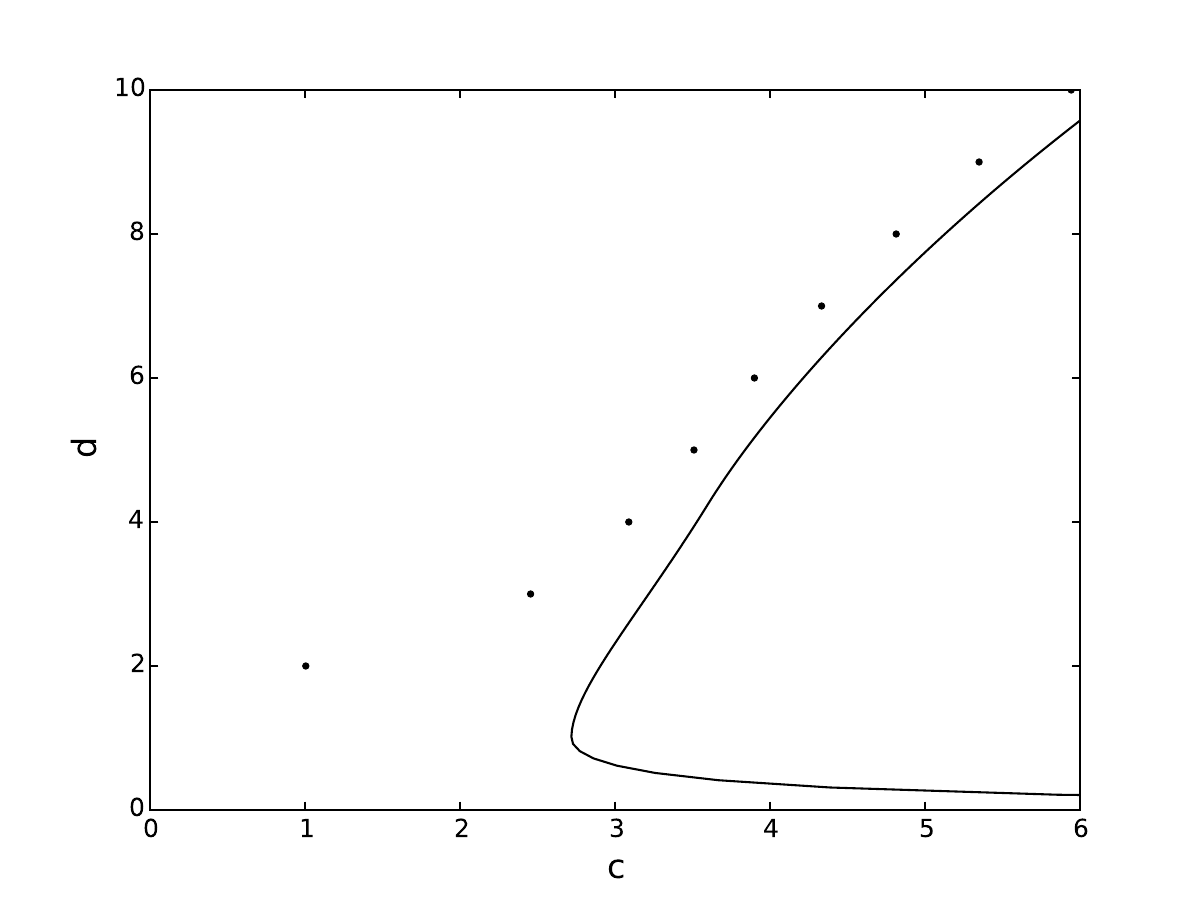}
\caption{Phase diagram of $(2,\max(r,2))$-core percolation on hypergraphs with
  Poisson vertex degree distributions. Black circles and black line represents the phase boundary of
$d$-uniform hypergraphs and hypergraphs with Poisson hyperedge
cardinality distribution, respectively. 
\label{fig2}} 
\end{figure}
\subsection{$K=2$ and $S=2$}

In this section we study the $(2,2)$-core. A similar definition of removable hyperedges was used in \cite{citeulike:11962060}, where the core obtained from the GLR procedure is used to study the vertex
cover problem in uniform hypergraphs.
%
%

In this case, Eq.~(\ref{2-core:eq1_2}) reduces to
\begin{equation}
\delta=\sum_{r=1}^{\infty} Q_\mathrm{h}(r) \alpha^{r-1}.\label{2-core:eq1_7}
\end{equation}
The relative size of the $(2,2)$-core can be calculated by considering the
probability that a randomly chosen vertex is connected to at least two
non-removable hyperedges and the probability that a degree-one vertex is
connected to a hyperedge with less than $(r-2)$ other degree-one vertices. This
results in  
\begin{equation}
s_\mathrm{2c}=\sum_{k=2}^{\infty} P_\mathrm{n}(k)
\sum_{l=2}^{k}
\binom{k}{l} 
( 1-\delta)^l
  \delta^{k-l}. 
\label{2-core:eq1_8}
\end{equation}


Eqs.~(\ref{2-core:eq1_5}) and (\ref{2-core:eq1_7}) have the same critical
point as Eqs.~(\ref{2-core:eq1_1}) and (\ref{2-core:eq1_2}). 
Therefore, for
$(2,2)$-core
we recover the result found in
the graph case that both the $(2,2)$-core and the GCC
emerge at the same critical point~\cite{RevModPhys.80.1275}. We can compute the critical point by combining Eq.~\ref{eq_f2c}, \ref{2-core:eq1_7} and \ref{chapter5_eq9} at $\alpha=\delta=1$,  obtaining
\begin{equation}
\frac{ d_2-d_1}{d_1}\frac{c^*_2 - c^*}{c^*}=1,\label{per:eq4}
\end{equation}
In this case the phase transition is
always continuous (see solid lines in
Fig.~\ref{graph1c}~c and d) and for the studied hyperedge cardinality and
vertex degree distributions we have $\eta=2$. 

As before, we assume that the moments of the vertex degree and hyperedge cardinality distributions do not diverge. Let us consider points around the critical point, such as, $\alpha=1-\zeta$ with $\zeta=0^+$  and  $c=c^*+\chi$ with $\chi=0^+$. We can  define the function from Eqs. \ref{eq_f2c} as a function of  $\alpha$ and $\chi$( i.e. $F_\mathrm{2c}(\alpha,c)$). By expanding $F_\mathrm{g}(\alpha,\chi)$ around the point $(\alpha,c)=(1,c^*)$, and combining it with result from Eq.~\ref{eq6_chapter5} at the critical point, we can rewrite  Eq.~\ref{eq_fg} as
\begin{equation}
\left.\frac{\partial^2 F_\mathrm{2c}(\alpha,c)}{\partial^2 \alpha}\right|_{(1,c^*)} \zeta^2-2\left.\frac{\partial^2 F_\mathrm{2c}(\alpha,c)}{\partial \alpha\partial c}\right|_{(1,c^*)} \chi\zeta+(\cdot \cdot \cdot)=0.
\label{giant:eq8}
\end{equation}
Note that for this case the Eq.~\ref{chapter5_eq9} is still valid.  We obtain 
\begin{equation}
\zeta\sim\chi.
\label{eq16}
\end{equation}
By combining this equations with the fact that $s_{2c}(\alpha^*)=0$ and the result from Eq.~\ref{2c_eq1}, we obtain
\begin{equation}
s_{2c}\sim  \chi^{2}.
\label{eq17}
\end{equation}

%
%
%

%


%

\newpage
\section{Core percolation on hypergraphs}
\label{section4}
In Sec.~\ref{dominant_set} we show the relation between the hyperedge covering problems and the dominating set problem Sec.~\ref{dominant_set}  We also show how our algorithm is a generalization of the greedy leaf removals procedure~\cite{Zhao2015} used to solve the dominant set problem in polynomial time, and that the core obtain trough our methodology is always smaller than produced by the greedy leaf removal. In Sec.~\ref{number_hyperedges} to Sec.~\ref{h_critical_exponent} we show the detailed derivations for the equations in the main text.
\subsection{Relation between hyperedge and vertex covering problems and the minimum dominating set problem}
\label{dominant_set}
\begin{figure}[!htb]
\centering
\includegraphics[width=0.8\textwidth]{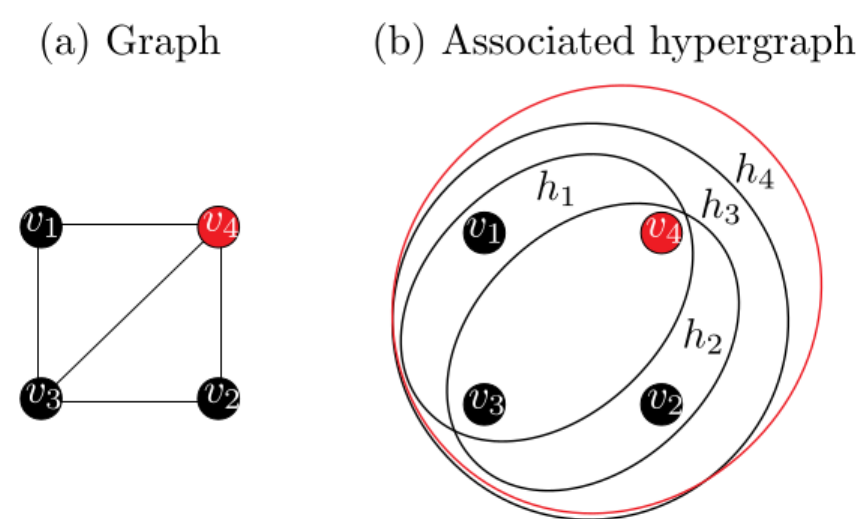}
\caption{A graph and the associated hypergraph. Hyperedge $h_i$ represents vertices observed by vertice $v_i$. Solving the hypergraph's minimum hyperedge cover set is equivalent to solve the graph's dominating set. The red vertex represents the solution of the minimum dominating set problem in  graph (a), the red edge the solution of the minimum hyperedge cover set problem.\label{Dominant}}
\end{figure}

The minimum dominating set (MDS) is the smallest set of vertices, $D$ of a graph $G$, so that every vertex in $G$ is adjacent to at least one vertex in $D$. The MDS can be mapped to a minimum hyperedge cover in a hypergraph composed by the same vertices as in $G$ and for each vertex $v_1$ there is an hyperedge, $h_1$, that contains $v_1$ and all neighbors of $v_1$ (see Figure~\ref{Dominant}). Therefore if we find the set of hyperedges $[v_i,...,v_j]$, that covers all vertices in this hypergraph, the dominanting set of graph $G$ is given by $[v_i,...,v_j]$. Our method  is a generalization of the method proposed in \cite{Zhao2015}. Note that our method doesn't focus onleaves, i.e., vertices with degree one. As shown in Figure~\ref{Dominant}, even given the fact that there is no one-degree vertices, our method can still solve the MDS exactly in polynomial time. 

{We emphasize that the greedy leaf removals rules described in~\cite{Zhao2015} are special cases of our method. Let us consider the first rule: if vertex $v_i$ is an unobserved leaf vertex (which has only a single neighbor, say $v_j$), then occupying $v_j$ but leaving $v_i$ empty must be an optimal strategy. And let us now consider the hypergraph with the same nodes as the original graph, and hyperedges $h_i$ that represent the vertices observed by node $v_i$. For a leaf $v_i$ the set of vertices that are covered by hyperedge $h_i$ is a subset or equal set to the ones covered by hyperedge $h_j$, i.e. $V_i\subseteq V_j$, then $h_i$ can be removed. After this, the hyperedges that cover $v_i$ are a subset of or equal to the set of hyperedges that cover $h_j$ ($H_i\subseteq H_j$), and a subset or equal set of hyperedges that cover the other neighbors of $v_j$, in the original network, $v_k$ ($H_i\subseteq H_k$). Then vertices $v_j$ and $v_k$ are removed. Note that all observed nodes are removed from the network, in our method, since they do not need to be observed anymore, but we still keep the hyperedges associated with them because they can still be used to cover other vertices. After this process the cardinality of hyperedge $h_j$ is one, meaning that $h_j$ is part of a minimum hyperedge cover set. In terms of dominating set it means that occupying node $j$ is an optimal strategy. Let us now consider the second rule : if vertex $v_i$ is an unobserved leaf vertex (which has only a single neighbor, say $v_j$), then occupying $v_j$ but leaving $v_i$ empty must be an optimal strategy, and if $v_i$ is an empty but observed vertex and at most one of its adjacent vertices is unobserved, then it must be an optimal strategy not to occupy $v_i$. We emphasize that in our hyperedge all observed vertices $v_k$ are automatically removed from the network. If all except at most one neighbor of vertex $i$ is unobserved,$v_j$, then $h_i$ can also be removed because all  vertices covered by hyperedge $v_i$ are also covered by vertex, $v_j$ ($H_i\subseteq H_j$). In terms of dominating set it means that not occupying vertice $v_i$ is an optimal strategy. 

Table~\ref{table1} shows the size of the cores associated with the dominating set, for the eleven networks analyzed in \cite{Zhao2015}. For most networks our method shows a considerable improvement, . For some of them, our method actually find no core left.}
\begin{table}[]
\centering
\setlength{\tabcolsep}{0.3cm}
\begin{tabular}{|c|c|c|c|c|c|}\hline
Networks   & $N$       & $M$       & $s_{\rm glr}$ & $s_{\rm d}$ \\\hline
RoadEU     & 1177    & 1417    & 0.260    & 0.167           \\\hline
PPI        & 2361    & 6646    & 0.007    & 0.000           \\\hline
Grid       & 4941    & 6594    & 0.122    & 0.056           \\\hline
IntNet1    & 6474    & 12572   & 0.001    & 0.000           \\\hline
Author     & 23133   & 93439   & 0.391    & 0.000           \\\hline
Citation   & 34546   & 420877  & 0.326    & 0.310           \\\hline
P2P        & 62586   & 147892  & 0.001    & 0.000           \\\hline
Friend     & 196591  & 950327  & 0.031    & 0.001           \\\hline
Email      & 265214  & 364481  & 0.002    & 0.000           \\\hline
Webpage    & 875713  & 4322051 & 0.185    & 0.004           \\\hline
RoadTX     & 1379917 & 1921660 & 0.406    & 0.342 \\\hline
\end{tabular}
\caption{Normalized size of the cores associated with  the minimum dominanting set problem for eleven real-world networka, using the greedy leaf removal procedure from\cite{Zhao2015}, $s_{\rm glr}$, and using our approach $s_{\rm d}$.  $N$ and $M$ are, respectively, the total number of nodes and links in the network. \label{table1}}
\end{table}

\subsection{Number of hyperedges contained in the core}
\label{number_hyperedges}
In the main text we characterize the size of the core as the number of vertices contained in the core, we can also compute the number of hyperedges in each core, $s'_v$ and $s'_h$. We can use the fact that the v$_{\mm{core}}$ of a hypergraph is h$_\mm{core}$ of the dual hypergraph. Thus, we can get  Eqs.  (1) (2) (7) and (8) in the main text by the following transformation
\begin{align}
&\alpha \rightarrow \delta\\
&\delta \rightarrow \beta\\
&\beta \rightarrow \epsilon\\
&\epsilon \rightarrow\alpha
 \end{align}
 on Eqs. (1) to (4) from the main text. Because the vertices of an hypergraph are hyperedges in the dual one, we obtain that  $s'_{\rm h}$  is given by
 \begin{equation}
s^{\rm h}_{\mathrm{h}}=\sum_{k=2}^{\infty} P_\mathrm{h}(r) \sum_{l=2}^{k}
\binom{k}{l} ( 1-\beta-\alpha)^l\alpha^{k-l},
\end{equation}
where $\alpha$, $\beta$ ,$\delta$ and $\gamma$ are obtained by solving Eqs. (1),(2) (7) and (8) in the main text. Using the same argument we can conclude that $s'_h$ is given by
 \begin{equation}
s^{\rm h}_{\mathrm{v}}=\sum_{k=2}^{\infty} P_\mathrm{h}(r) \sum_{l=2}^{k}
\binom{k}{l} ( 1-\alpha-\beta)^l\beta^{k-l},
\end{equation}
where $\alpha$, $\beta$ ,$\delta$ and $\gamma$ are obtained by solving Eqs. (1),(2) (7) and (8) in the main text.
\subsection{$v_{ \rm core}$ for uniform hypergraphs - critical point}
There is a simple relation between $v_{\rm core}$ critical point and the mean cardinality $d_1$ for uniform hypergraphs with $d_1>1$. Let us start by rewriting Eq.~1 in the main text as
\begin{equation}
F_\mathrm{c1}(\alpha,c)=0
\label{eq18}
\end{equation}
where we define
\begin{equation}
F_\mathrm{c1}(\alpha,c)\equiv \sum_{k=1}^{\infty} Q_\mathrm{n}(k) \epsilon^{k-1}-\alpha.
\label{eq19}
\end{equation}
For  uniform-Poisson hypergraph Eqs (1), (2) and (7) and (8) in the main text can be reduced to
\begin{align}
x&=e^{-c(d_1-1)y}\label{eq20a}\\
y&=e^{-c(d_1-1)x}\label{eq20b},
\end{align}
where we define $x\equiv \alpha^{r-1}$  and $y\equiv (1-\beta)^{r-1}$. The function  defined in Eq.~\ref{eq19} can be written as a function of $x$ and $c$,
\begin{equation}
F_\mathrm{c1}(x,c)\equiv \exp{\left(-c~(d_1-1)~e^{-c~(d_1-1)~x}\right)}-x.
\label{eq21}
\end{equation}
The critical point is given by $\alpha^*=(c^*~(d_1-1))^{1/(1-d_1)}$ and $c^*=e/(d_1-1)$.
\subsection{$v_{ \rm core}$ - critical exponent}
Eqs. 1 and 2, 7 and 8  in the main text can be written as
\begin{align}
\alpha&=A(1-z)\\
z&=B(h)\\
h&=A(1-\delta)\\
\delta&=B(\alpha)
\end{align}
where we define $z\equiv 1-\epsilon$, $h\equiv1-\beta$, $A(1-z)\equiv\sum_{k=1}^{\infty} Q_\mathrm{n}(k) (1-z)^{k-1}$ and $\left.B(h)\equiv\sum_{r=2}^{+\infty}  Q_r(r)h^{r-1}\right.$. $ F_\mathrm{v}(\alpha,c)$ can be written as
\begin{equation}
F_\mathrm{v}(\alpha,c)=H(H(\alpha,c),c)-\alpha,
\end{equation}
where we define
\begin{equation}
H(\alpha,c)\equiv A(1-B(\alpha),c1)\label{H}.
\end{equation}
This function has the same form for the graph case,  (see Eq.~S60 from \cite{PhysRevLett.109.205703}). The only difference is the way  we define the function $H(\alpha,c)$. Nevertheless it obeys the same relation, and therefore has the same critical exponent $\zeta_1=1$, i.e:
\begin{equation}
\left.\partiald{H(\alpha,c)}{\alpha}\right|_{(\alpha,c)=(\alpha^*,c^*)}=-1,\label{relation1}
\end{equation}
\begin{equation}
\left.\partiald{F_\mathrm{h}(\alpha,c)}{c}\right|_{(\alpha,c)=(\alpha^*,c^*)}=0,
\end{equation}
\begin{equation}
\left.\partialdt{F_\mathrm{h}(\alpha,c)}{\alpha}\right|_{(\alpha,c)=(\alpha^*,c^*)}=0.
\end{equation} 
From an expansion of $F_\mathrm{h}$ as a function of $(c-c^*)$, for $c-c^*=0^+$, it follows that
\begin{equation}
\alpha-\alpha^*\sim  (c-c^*)^{1/2},
\end{equation}
\begin{equation}
\beta-\beta^*\sim  (c-c^*)^{1/2},
\end{equation}
\begin{equation}
1-\beta-\alpha\sim  (c-c^*)^{1/2}\label{relation2},
\end{equation}
and
\begin{equation}
s^{\rm v}_{v}\sim  (c-c^*),
\end{equation}
witch is equivalent to Eq.~(6) in the main text, for $\zeta_1=1$. 
\subsection{$h_{\rm core}$ - critical exponent}
\label{h_critical_exponent}
Eqs.~(1) to (4) in the main text can be written as
\begin{align}
\alpha&=A(\epsilon)\\
\epsilon&=B(1-u)\\
u&=A(v)\\
v&=B(1-\alpha)
\end{align}
where we define $u\equiv 1-\beta$ and $v\equiv 1-\delta$. The function $F_\mathrm{c2}(\alpha,c)\equiv A(\epsilon,c)-\alpha$ can be written as
\begin{equation}
F_\mathrm{h}(\alpha,c)=G(G(\alpha,c),c)-\alpha,
\end{equation}
where we define
\begin{equation}
G(\alpha,c)\equiv A(B(1-u),c1).
\end{equation}
This function has the same form as Eqs.~\ref{H}. The only difference is how we define the function $G(\alpha,c)$. Nevertheless it obeys  the same relations as Eqs.~\ref{relation1} to \ref{relation2}, thus,
\begin{equation}
s^{\rm v}_{\rm h}\sim  (c-c^*)^{\zeta_2},
\end{equation}
with $\zeta_2=\zeta_1=1$.


\clearpage
\section{Simulations}
\label{section5}
We compare our analytical  calculations with  random hypergraphs generated in the following ways.
\subsubsection{Poisson-Uniform hypergraphs}
We first  generate random hypergraphs where all hyperedges have the same cardinality $d$ and the vertex hyperdegrees follow  a Possion distribution with average value $c_1$.
 We can generate this hypergraphs by creating $N$ nodes and $L$ hyperedges of cardinality $d$. We then fill the hyperedges with random selected vertices from our list of nodes. In the end
 \begin{equation}
 c=dL/N.
 \end{equation}
\subsubsection{Poisson-Poisson hypergraphs}
In the Poisson-Poisson hypergraphs we generate random hypergraphs where and both the vertex degrees and hyperedge cardinalities follow a Possion distribution with average  $c$ (or $d$), respectively. 
 We can generate this hypergraphs by creating $N$ vertices and $L$ hyperedges. We then select random vertices and add it to a random selected hyperedge until the average hyperdegree of the network is
  \begin{equation}
 c=d \times L/N.
 \end{equation}
 \subsection{Results}
\begin{figure}[!htb]
\centering
\includegraphics[width=1 \textwidth]{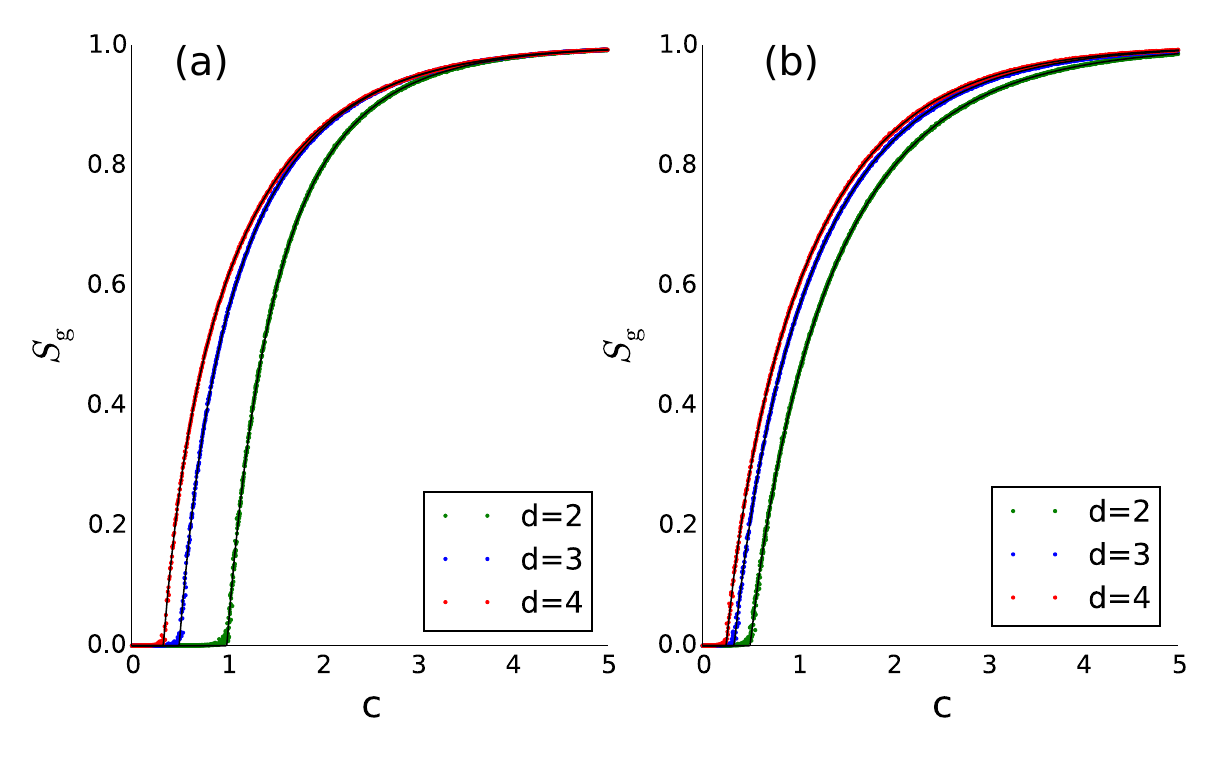}
\caption{The relative size of the giant connected component, $s_\mm{g}$, for hypergraphs with Poisson vertex degree distributions and mean degree $c$. Black solid lines: theoretical results. Dots: numerical results from hypergraphs with $10^4$ vertices. (a) $d$-uniform hypergraphs where all hyperdege have the same cardinality $d$. (b) the hyperedge cardinality follows a Poisson distribution with average cardinality $d$.\label{fig1}}
\end{figure}

 Figures \ref{fig1} to \ref{fig5} show that numerical simulations agree well with  the analytical calculations. In the simulations we used hypergraphs of size $10^4$ vertices. 
\begin{figure}[!htb]
\centering
\includegraphics[width=1 \textwidth]{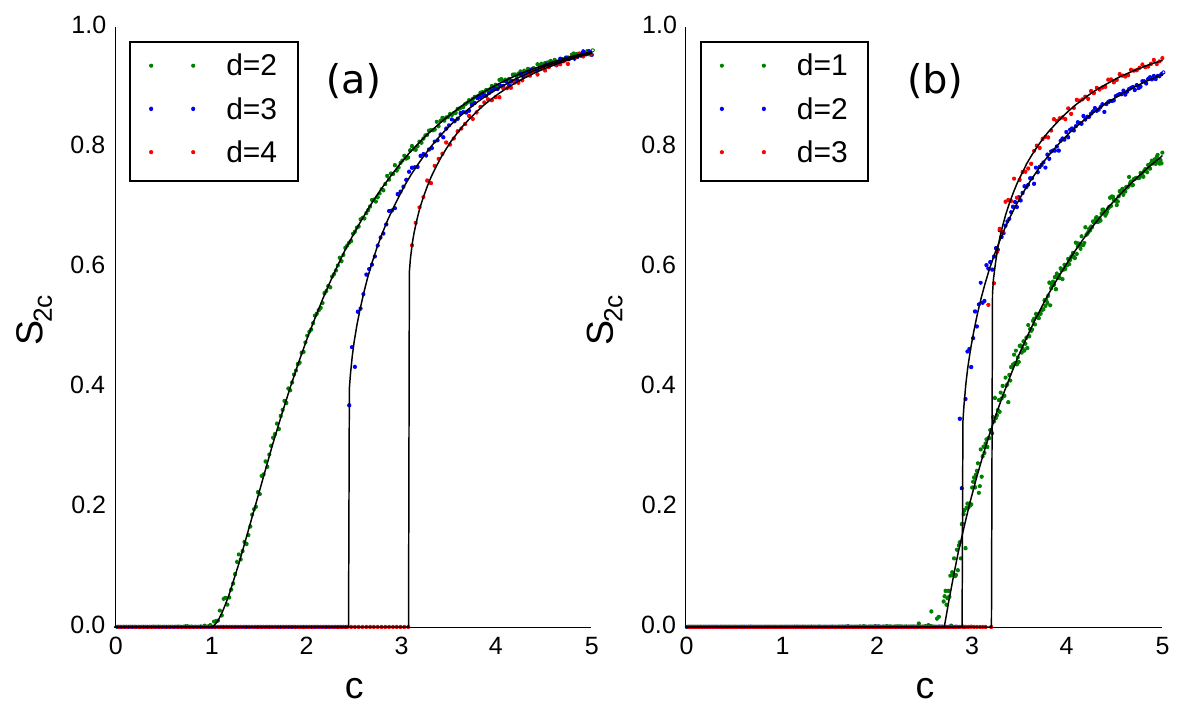}
\caption{(a) The relative size of the $(2,S)$-core with $S=\max\left(r-1,2\right)$ for hypergraphs with Poisson vertex degree distributions and mean degree $c$. Black solid lines: theoretical results. Dots: numerical results from hypergraphs with $10^4$ vertices. (a) $d$-uniform hypergraphs where all hyperdege have the same cardinality $d$. (b) the hyperedge cardinality follows a Poisson distribution with average cardinality $d$. \label{fig2}}
\end{figure}
\begin{figure}[!htb]
\centering
\includegraphics[width=1 \textwidth]{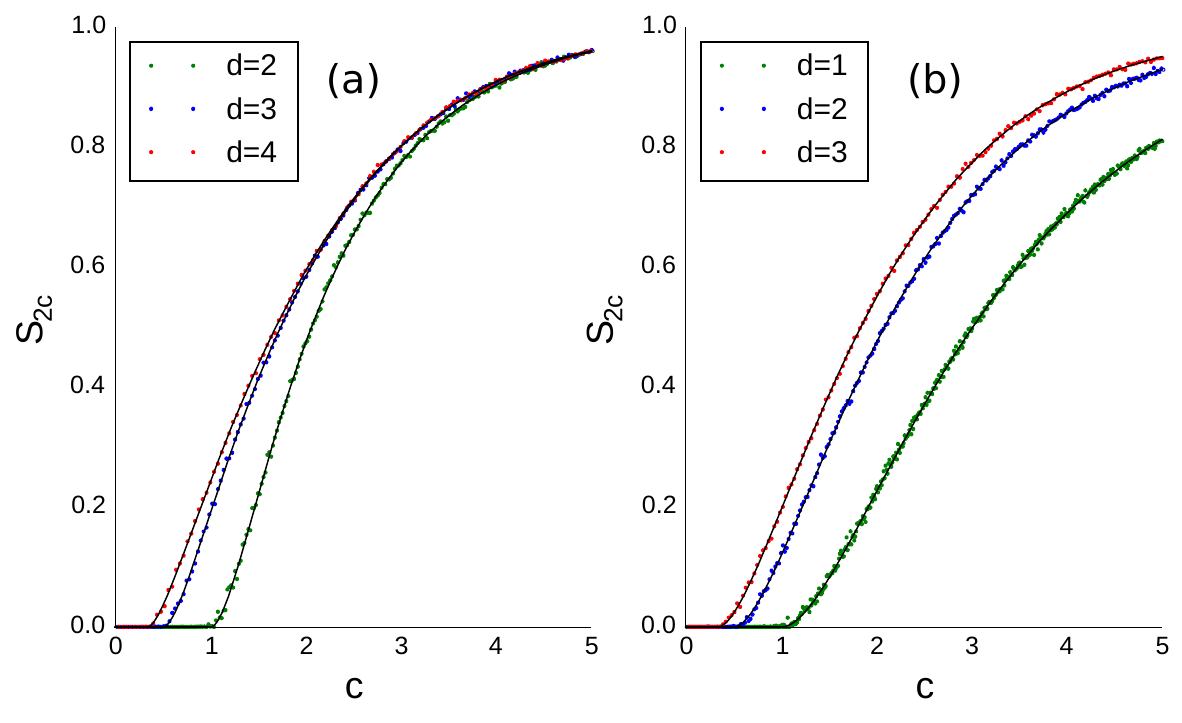}
\caption{The relative size of the $(2,S)$-core with $S=2$ for hypergraphs with Poisson vertex degree distributions and mean degree $c$. Black solid lines: theoretical results. Dots: numerical results from hypergraphs with $10^4$ vertices. (a) $d$-uniform hypergraphs where all hyperdege have the same cardinality $d$. (b) the hyperedge cardinality follows a Poisson distribution with average cardinality $d$.  \label{fig3}}
\end{figure}

\begin{figure}[!htb]
\centering
\includegraphics[width=1 \textwidth]{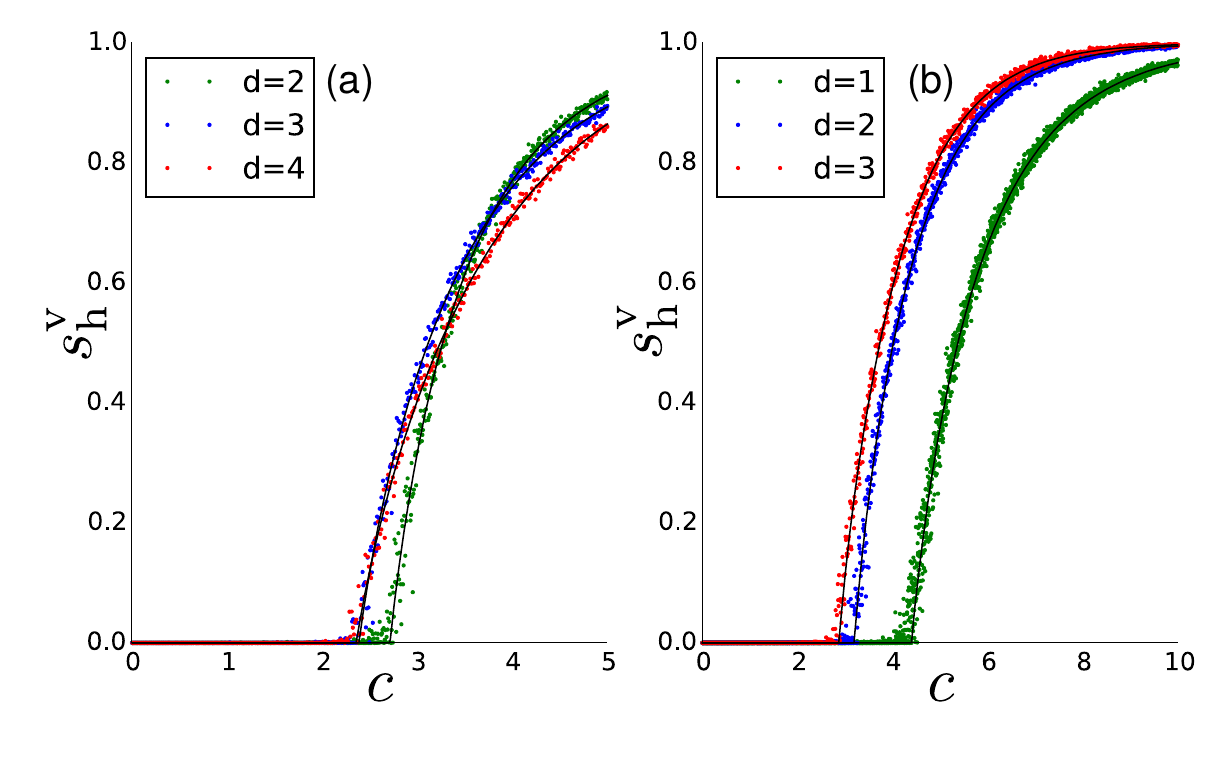}
\caption{\textbf{$h_{\rm core}$ obtained from simulations.} The relative size of the h-core for hypergraphs with Poisson vertex degree distributions and mean degree $c$. Black solid lines: theoretical results. Dots: numerical results from hypergraphs with $10^5$ vertices. (a) $d$-uniform hypergraphs where all hyperdege have the same cardinality $d$. (b) the hyperedge cardinality follows a Poisson distribution with average cardinality $d$.\label{fig4}}
\end{figure}
\begin{figure}[!htb]
\centering
\includegraphics[width=1 \textwidth]{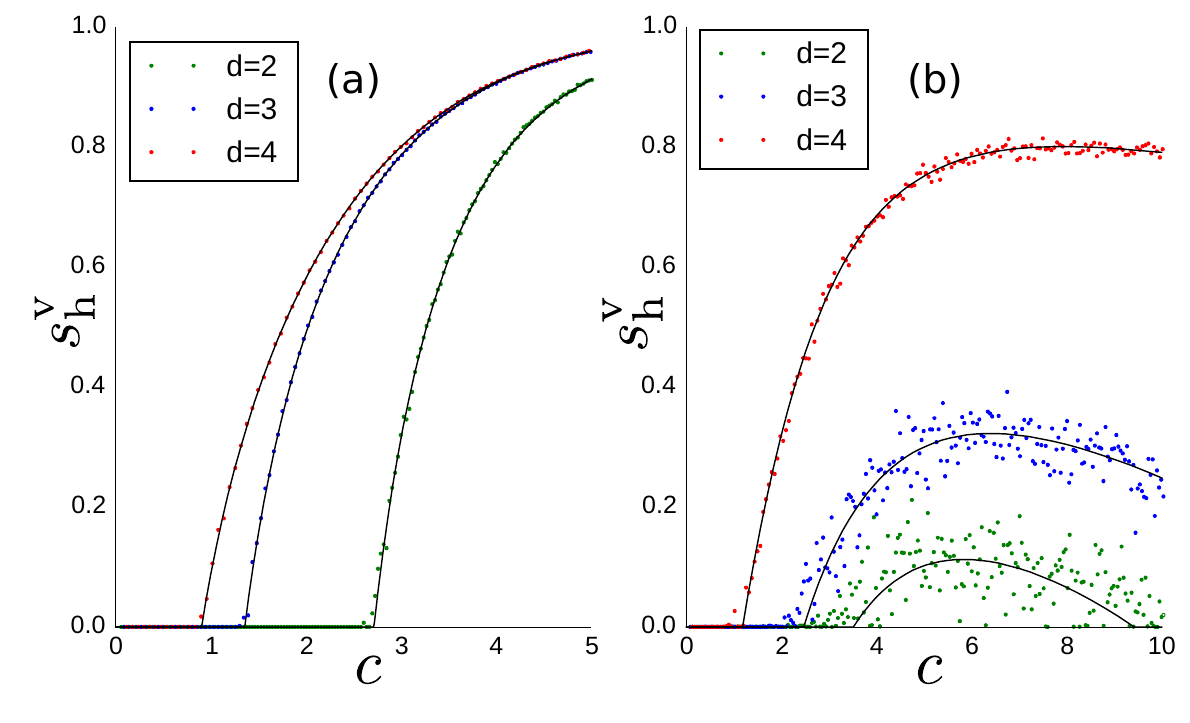}
\caption{\textbf{$v_{\rm core}$ obtained from simulations.} The relative size of the v-core for hypergraphs with Poisson vertex degree distributions and mean degree $c$. Black solid lines: theoretical results. Dots: numerical results from hypergraphs with $10^5$ vertices. (a) $d$-uniform hypergraphs where all hyperdege have the same cardinality $d$. (b) the hyperedge cardinality follows a Poisson distribution with average cardinality $d$. The large fluctuation for  $d=2$ and $3$ are due to the relatively large number of hypereges with cardinalty $1$, and the chance that if a edge is connected to a  hypereges with cardinalty $1$ is automatically removed.\label{fig5}}
\end{figure}

\clearpage

\section{Contributions and Author Information}
Contributions. Y.-Y.L. and H.-J.Z conceived the project. All authors designed the research. B.C.C. performed all the analytical calculations and numerical simulations.
~All authors analyzed the results. B.C.C. and Y.-Y.L. wrote the manuscript. H.-J.Z. edited the manuscript.

Author Information. The authors declare no competing financial interests. Correspondence and requests for materials should be addressed to Y.-Y.L. (yyl@channing.harvard.edu), or B.C.C. (coutinho.b@husky.neu.ed).
\clearpage
\bibliographystyle{unsrt}
\bibliography{beta}
\clearpage